\documentclass[11pt]{article}
\usepackage{amssymb}
 \linethickness{0.4pt}
\begin{document}
\newtheorem {proposition}{Proposition}[section]
\newtheorem{lemma}{Lemma}[section]
\newtheorem{theorem}{Theorem}[section]
\newtheorem{corollary}{Corollary}[section]
\begin{flushright}
                IFT UwB /14/2000 \\

\end{flushright}

\bigskip
\bigskip
\begin{center}
{\large\bf THE INTEGRABLE MULTIBOSON SYSTEMS\\ AND\\ ORTHOGONAL
POLYNOMIALS }\footnote{Supported in part by KBN grant 2 PO3 A 012
19

E-mail: *aodzijew@labfiz.uwb.edu.pl, **horowski@alpha.uwb.edu.pl,
***tereszk@labfiz.uwb.edu.pl}
\end{center}
\bigskip
\bigskip
\begin{center}
{\bf Anatol Odzijewicz*, Maciej Horowski**, Agnieszka
Tereszkiewicz*** }

\end{center}
\bigskip
\bigskip
\begin{center}
{Institute of Theoretical Physics\\ University in Bia{\l}ystok
\\ul. Lipowa 41, 15-424 Bia{\l}ystok, Poland}
\end{center}
\bigskip\bigskip

\begin{abstract}
\noindent The strict relation between some class of multiboson
hamiltonian systems and the corresponding class of orthogonal
polynomials is established. The correspondence is used effectively
to integrate the systems. As an explicit example we integrate the
class of multiboson systems corresponding to q-Hahn polynomials.
\end{abstract}
\vspace{1.5cm}
\thispagestyle{empty}

\section*{Introduction}
The results of this paper give an effective tools to integrate
some class of quantum physical systems. Such systems play a
prominent role in quantum many-body physics, nuclear physics as
well as quantum optics. More information about these models and
their physical content can be found in many papers and monographs
on this subject and among others one may consult \cite{A-I},
\cite{J1}, \cite{J2}, \cite{Kar1}, \cite{Kar2}, \cite{M-G},
\cite{Odz}.

This paper is devoted to a detailed study of the mathematical
structures underlying the multi-boson hamiltonian systems, whose
dynamics is generated by the operators of the general form given
in (\ref{Aa}).

In the first section, it is shown, that the reduction procedure
applied to the multiboson hamiltonian system leads to some
interesting operator algebras $\cal A_R$ (parametrized by a
structural function $\cal R$) which generalize in a natural way
Heisenberg algebra as well as $sl_q(2)$ and $SU_q(2)$ quantum
algebras. It was demonstrated  \cite{Odz} that the algebras $\cal
A_R$ are on the one hand side strictly related to the integration
of the quantum systems. Their description in terms of the coherent
states leads on the other hand to the connection of these algebras
with the theory of basic hypergeomeric series and orthogonal
polynomials.

In Section 2 we establish the explicit relation between the
spectral realization of the algebra $\cal A_R$ in the space
$L^2\left({\Bbb R},d\sigma\right)$ and the representation in the
space of holomorphic function.

According to the spectral theorem, the problem of integration of
the dynamical system is equivalent to the construction of the
spectral measure $d\sigma$ for the corresponding Hamiltonian.
Under some assumptions on its form (see Section 3) the problem of
this construction  can be explicitly solved within the framework
of q-Hahn's polynomials theory. The measure is then uniquely
determined by the solution of $q$-difference Pearson equation.

For the sake of completeness some elementary facts from
$q$-analysis are presented in the Appendix A. The Appendix B
contains the proofs of some fundamental properties of q-Hahn's
polynomials.

\section{The multi-boson systems}
\renewcommand{\theequation}{1.\arabic{equation}}
\setcounter{equation}{0} This section is devoted to the detailed
analysis of the symmetry properties of ($N+1$)-boson systems. The
dynamics of these systems is assumed to be governed by the
Hamiltonian operator of the form:
\begin{equation}
\label{Aa}
\begin{array}{c}
H=h_0\left( a_0^{*}a_0,\ldots ,a_N^{*}a_N\right) +g_0\left(
a_0^{*}a_0,\ldots ,a_N^{*}a_N\right) a_0^{k_0}\ldots a_N^{k_N}+ \\
+a_0^{-k_0}\ldots a_N^{-k_N}\overline{g}_0\left( a_0^{*}a_0,\ldots
,a_N^{*}a_N\right),
\end{array}
\end{equation}
where $a_0,\ldots ,a_N$ and $a_0^{*},\ldots ,a_N^{*}$ are bosonic
annihilation and respectively creation operators with standard
Heisenberg commutation relations:
\begin{equation}
\label{Ab}
\left[ a_i,a_j^{*}\right] =\delta _{ij},\quad \left[
a_i,a_j\right] =0,\quad \left[ a_i^{*},a_j^{*}\right] =0.
\end{equation}
The following notational convention is assumed in (\ref{Aa})
\begin{equation}
\label{Ac} a_i^{k_i}=\left\{
\begin{array}{l}
a_i^{k_i}\quad \;\quad \quad \quad\;\;\; for\;k_i> 0\\ 1\quad
\;\;\;\;\quad \quad \quad\;\;\; for k_i=0\\a_i^{k_i}=\left(
a_i^{*}\right) ^{-k_i}\;\;for\;k_i<0.
\end{array}
\right.
\end{equation}
The monomial
\begin{equation}
\label{Ad}
a_0^{k_0}\ldots a_N^{k_N},\quad k_0,\ldots ,k_N\in
{\Bbb {Z}}
\end{equation}
can be thought of as an operator which describes the subsequent
creation and annihilation of the clusters of the bosonic modes.\\
The operator
\begin{equation}
\label{Ae}
g_0\left( a_0^{*}a_0,\ldots ,a_N^{*}a_N\right)
\end{equation}
is a kind of generalization of the coupling constant. The coupling
constant is replaced in our case by a function depending on the
occupation number operators of the bosonic modes. The operator
\begin{equation}
\label{Af}
 h_0\left( a_0^{*}a_0,\ldots ,a_N^{*}a_N\right)
\end{equation}
can always be chosen as a free Hamiltonian being a weighted sum of
the occupation number operators of elementary modes $a_0^{*}a_0,\ldots ,a_N^{*}a_N$:%
\begin{equation}
\label{Ag} h_0^{free}= \omega _0a_0^{*}a_0+\ldots +\omega
_Na_N^{*}a_N.
\end{equation}

The considerations at this paper will be by no means restricted to
this free case however. The Hamiltonian under consideration
(\ref{Aa}) is an elementary ingredient of the most general
Hamiltonian operator
\begin{equation}
\label{Ah}
H=\sum_{k_0,\ldots ,k_N\in {\Bbb {Z}}}g_{k_0\ldots
k_N}\left( a_0^{*}a_0,\ldots ,a_N^{*}a_N\right) a_0^{k_0}\ldots
a_N^{k_N},
\end{equation}
with the functions $g_{k_0\ldots k_N}$ connected by the following
conjugation rule.
\begin{equation}
\label{Ai}
\left[ g_{k_0\ldots k_N}\left( a_0^{*}a_0,\ldots
,a_N^{*}a_N\right) \right] ^{*}=g_{-k_0\ldots -k_N}\left(
a_0^{*}a_0-k_0,\ldots ,a_N^{*}a_N-k_N\right) .
\end{equation}
The class of model hamiltonians (\ref{Aa}) corresponds to many
important quantum physical systems. Their dynamics is generated by
specific operators of the form (\ref{Aa}) \cite{J1}, \cite{J2},
\cite{Kar1}, \cite{Kar2}, \cite{A-I}. For this reason the analysis
of the system (\ref{Aa}) seems to be important and may shed new
light on the unsolved problems of quantum physics.

In order to analyze the quantum system described by the
Hamiltonian (\ref{Aa}), it is convenient  to introduce the
following operators:
\begin{equation}
\label{Aj}
 A:=g_0\left( a_0^{*}a_0,\ldots ,a_N^{*}a_N\right)
a_0^{k_0}\ldots a_N^{k_N}.
\end{equation}
and
\begin{equation} \label{Ak} A_i=A_i^{*}:={\sum_{j=0}^N}\alpha
_{ij}a_j^{*}a_j\ ,
\end{equation}
where\ $i=0,1,2,\ldots ,N$. One assumes that real $\left(
N+1\right) \times \left( N+1\right) $-matrix\\$\alpha =\left(
\alpha _{ij}\right) $ satisfies the conditions
\begin{eqnarray}
\label{Al}
 \det \alpha &\neq& 0 \\ \label{All}
 {\sum_{j=0}^N}\alpha _{ij}k_j &=& \delta _{i0}.
\end{eqnarray}
The operators $A_0,A_1,\ldots ,A_N,A$ and $A^{*}$ do satisfy the
following commutation relations
\begin{equation}
\left[ A_0,A\right] =-A,\quad \left[ A_0,A^{*}\right] =A^{*}
\label{Am}
\end{equation}
\begin{equation}
\left[ A,A_i\right] =0=\left[ A^{*},A_i\right]  \label{An}
\end{equation}
for $i=1,\ldots N$, and
\begin{equation}
\left[ A_i,A_j\right] =0  \label{Ao}
\end{equation}
for $i,j=0,1,\ldots ,N$.

One has in addition
\begin{eqnarray}
A^{*}A &=&\left| g_0\left( a_0^{*}a_0-k_0,\ldots
,a_N^{*}a_N-k_N\right) \right| ^2{\cal P}_{k_0}\left(
a_0^{*}a_0-k_0\right) \ldots {\cal P}_{k_N}\left(
a_N^{*}a_N-k_N\right) \label{Ap} \\ AA^{*} &=&\left| g_0\left(
a_0^{*}a_0,\ldots ,a_N^{*}a_N\right) \right| ^2{\cal
P}_{k_0}\left( a_0^{*}a_0\right) \ldots {\cal P}_{k_N}\left(
a_N^{*}a_N\right) , \label{Ar}
\end{eqnarray}
where ${\cal P}_{k_0}\left( a_0^{*}a_0\right) ,\ldots, {\cal
P}_{k_N}\left( a_N^{*}a_N\right) $ are polynomials:
\begin{equation}
{\cal P}_k\left( a^{*}a\right) :=a^ka^{-k}=\left\{
\begin{array}{l}
a^k\left( a^{*}\right) ^k=\left( a^{*}a+1\right) \ldots \left(
a^{*}a+k\right) \quad \quad \quad \quad \ \quad \,for\;k>0 \\
1\quad \;\;\;\quad \quad \quad \quad \quad \quad \quad \quad \quad
\quad \quad \quad \quad \,\,\,\quad \quad \quad \quad for\;k=0 \\
\left( a^{*}\right) ^{-k}a^{-k}=a^{*}a\left( a^{*}a-1\right)
\ldots \left( a^{*}a-k+1\right) \quad for\;k<0.
\end{array}
\right.  \label{As}
\end{equation}

The operators $A^{*}A$ and $AA^{*}$ are diagonal in the standard
Fock basis
\begin{equation}
\left| n_0,n_1,\ldots ,n_N\right\rangle =\frac 1{\sqrt{n_0!\ldots n_N!}%
}\left( a_0^{*}\right) ^{n_0}\ldots \left( a_N^{*}\right)
^{n_N}\left| 0\right\rangle ,  \label{Aw}
\end{equation}
where $\left( n_0,n_1,\ldots ,n_N\right) \in {\Bbb {Z}}_{+}^{N+1}:{\Bbb =}%
\left( {\Bbb Z}_{+}\cup \left\{ 0\right\} \right) \times \ldots
\times \left( {\Bbb Z}_{+}\cup \left\{ 0\right\} \right) $ ($N+1$
times).

Let us note that operators  $A_0,\ldots ,A_N$ are unbounded.
Whether $A$ and $A^{*}$ are bounded or not depends on the choice
of the structural function $g_0$. All of them are defined on the
common domain $D$ spanned by finite linear combinations
\begin{equation}
\left| v\right\rangle ={\sum_{\left( i_0,i_1,\ldots ,i_N\right) \in F}}%
c_{i_0,i_1,\ldots ,i_N}\left| n _{i_0},n_{i_1},\ldots, n_{i_N}
\right\rangle \label{Auu}
\end{equation}
of the Fock basis elements, where $F$ is some finite set of multi
indices.

Identifying the 1-dimensional spaces ${\Bbb C}\left|
n_0,n_1,\ldots ,n_N\right\rangle $ with the elements of ${\Bbb
Z}_{+}^{N+1}$ we obtain the action of $A$ and $A^{*}$ (and their
natural powers) on ${\Bbb Z}_{+}^{N+1}$. It is easy to see that
the orbits of these actions are located on one dimensional lines
which are parallel to the vector $\left( k_0,k_1,\ldots
,k_N\right) \in {\Bbb Z}^{N+1}$.

If the function $g_0$ of ( \ref{Aa}) is regular and nonvanishing
in all points of ${\Bbb Z}_{+}^{N+1}$, we can make some simple but
useful observations. The first one is that if $\frac{k_i}{k_j}<0$
for some $i,j\in \left\{ 0,1,\ldots ,N\right\} $ then for any
element $\left| n_0,n_1,\ldots ,n_N\right\rangle $ of the Fock
basis there exists $M\in {\Bbb {N}}$ such that
\begin{equation}
A^M\left| n_0,n_1,\ldots ,n_N\right\rangle =0\quad and\quad \left(
A^{*}\right) ^M\left| n_0,n_1,\ldots ,n_N\right\rangle =0.
\label{Aww}
\end{equation}
This means that orbits of $A$ and $A^*$ in ${\Bbb {Z}}^{N+1}_+$
are finite. In the opposite case the orbits contain infinitely
many points.

 The second observation is that $A\left| n_0,n_1,\ldots
,n_N\right\rangle =0$ if and only if there exists $i\in \left\{
0,1, \ldots ,N \right\}$ such that $k_i>0$ and
$n_i\in\left\{0,1,\ldots,k_i-1\right\}$. Hence each orbit of $A$
and $A^*$ in ${\Bbb {Z}}^{N+1}_+$ has exactly one vacuum (the
 point anihilated by $A$). It is then natural to introduce the
 parametrization of the Fock basis which is in agreement with the above orbit
decomposition of ${\Bbb {Z}_+^{N+1}}$.

Replacing the occupation number operators $ a_0^{*}a_0,\ldots
,a_N^{*}a_N $ by the operators $ A_0,A_1,\ldots ,A_N $ in
(\ref{Ap}) and (\ref{Ar}) one obtains:
\begin{eqnarray}
A^{*}A &=&{\cal G}\left( A_0-1,A_1,\ldots ,A_N\right)  \label{Ass}
\\ AA^{*} &=&{\cal G}\left( A_0,A_1,\ldots ,A_N\right)  \label{At}
\end{eqnarray}
with the function ${\cal G}$ uniquely determined by $g_0$,
polynomials ${\cal P}_{k_0},\ldots ,{\cal P}_{k_N}$ and the linear
map (\ref{Ak}).

The Hamiltonian (\ref{Aa}) can be reexpressed in terms of
(\ref{Aj}) and (\ref{Ak}) in the following form:
\begin{equation}
H=H_0\left( A_0,A_1,\ldots ,A_N\right) +A+A^{*}  \label{Au}
\end{equation}
It is clear that it admits $N$ commuting integrals of motion
$A_1,\ldots ,A_N$: $\left[ A_i , H\right]=0$ for $i=1,\ldots,N$.
They commute with the operator\ $A_0$ too. This maximal system of
commutative observables is diagonalized in the Fock basis and the
eigenvalues of $A_0,A_1,\ldots ,A_N$ on $ \left| n_0,n_1,\ldots
,n_N\right\rangle $ are given by
\begin{equation}
\lambda _i={\sum_{j=0}^N}\alpha _{ij}n_j\quad i=0,1,\ldots ,N.
\label{Az}
\end{equation}
The eigenvalues $\left( \lambda _0,\lambda _1,\ldots ,\lambda
_N\right) $ form a discrete convex cone $\ \Lambda
_{+}\subset{\Bbb {R}}^{N+1}$. It is spanned by the columns of the
matrix $\left( \alpha _{ij}\right) $ with entries from $\left(
{\Bbb {Z}_{+}}\cup \left\{ 0\right\} \right) $: $\Lambda
_{+}=\alpha \left( {\Bbb Z}_{+}^{N+1}\right) $. The sequences
$\left( \lambda _0,\lambda _1,\ldots ,\lambda _N\right) \in
\Lambda _{+}$ will be used as a new parametrization $\left\{
\left| \lambda _0,\lambda _1,\ldots ,\lambda _N\right\rangle
\right\} $ of the  Fock basis elements.

In order to integrate the system (\ref{Aa}) one can reduce it to
the eigen subspaces ${\cal H}_{\lambda _1\ldots \lambda _N}\subset
{\cal H}$ spanned by the eigenvectors $\left| \lambda _0,\lambda
_1,\ldots ,\lambda _N\right\rangle $ with fixed $\lambda _1,\ldots
,\lambda _N$. This subspace is invariant with the respect to the
algebra ${\cal A}_{red}\ $, which is generated by the operators
$A,A^{*},$ $A_0\ $. These operators do satisfy the following
relations (\ref{Ass}),(\ref{At}),(\ref{Am}):
\begin{eqnarray}
\left[ A_0,A\right] &=&-A,\quad \left[ A_0,A^{*}\right] =A^{*},
\label{Av}
\\
A^{*}A &=&{\cal G}\left( A_0-1,\lambda _1,\ldots ,\lambda
_N\right) , \label{Ax} \\ AA^{*} &=&{\cal G}\left( A_0,\lambda
_1,\ldots ,\lambda _N\right) . \label{Ay}
\end{eqnarray}
Hence one can conclude that the problem of integration of the
system (\ref {Aa}) amounts to the integration of the system
described by the reduced Hamiltonian:
\begin{equation}
H_{red}=H_0\left( A_0,\lambda _1,\ldots ,\lambda _N\right)
+A+A^{*} \label{Azz}
\end{equation}
being an element of the algebra ${\cal A}_{red}$.

The orthonormal basis of the Hilbert subspace ${\cal H}_{\lambda
_1\ldots \lambda _N}$ is formed by the vectors $\left|\; \lambda
_0,\lambda _1,\ldots ,\lambda _N\;\right\rangle $ with $\lambda_0$
such that $\left( \lambda _0,\lambda _1,\ldots ,\lambda _N\right)
\in \Lambda _{+}$. From (\ref{Av}-\ref{Ay}) it follows that
\begin{eqnarray} A_0\left| \lambda _0,\lambda _1,\ldots
,\lambda _N\right\rangle &=&\lambda _0\left| \lambda _0,\lambda
_1,\ldots ,\lambda _N\right\rangle  \label{A} \\
A\left| \lambda _0,\lambda _1,\ldots ,\lambda _N\right\rangle &=&\sqrt{{\cal %
G}\left( \lambda _0-1,\ldots ,\lambda _N\right) }\left| \lambda
_0-1,\lambda_1,\ldots ,\lambda _N\right\rangle  \label{Aaa} \\
A^{*}\left| \lambda _0,\lambda _1,\ldots ,\lambda _N\right\rangle
&=&\sqrt {{\cal G}\left( \lambda _0,\lambda _1,\ldots ,\lambda
_N\right) }\left| \lambda _0+1,\lambda_1,\ldots ,\lambda
_N\right\rangle . \label{AA}
\end{eqnarray}
Let us note here that if $\left( \lambda _0,\lambda _1,\ldots
,\lambda _N\right) \in \Lambda _{+}$ then either $\left( \lambda
_0-1,\ldots ,\lambda _N\right) \in \Lambda _{+}$\\\quad ($\left(
\lambda _0+1,\ldots ,\lambda _N\right) \in \Lambda _{+} $) or
\begin{equation}
A\left| \lambda _0,\lambda _1,\ldots ,\lambda _N\right\rangle
=0\quad \left( A^{*}\left| \lambda _0,\lambda _1,\ldots ,\lambda
_N\right\rangle =0\right) . \label{Aab}
\end{equation}
Because of (\ref{Ax}-\ref{Ay}) the conditions (\ref{Aab}) are
equivalent to
\begin{equation}
{\cal G}\left( \lambda _0-1,\lambda _1,\ldots ,\lambda _N\right)
=0\quad \left( {\cal G}\left( \lambda _0,\lambda _1,\ldots
,\lambda _N\right) =0\right).  \label{Aac}
\end{equation}

Since $\Lambda _{+}$ is a discrete convex cone we can easily see
that the representation of ${\cal A}_{red}$ in ${\cal H}_{\lambda
_1\ldots \lambda _N} $ splits into irreducible components. These
components are generated out of vacuum (or antivacum) states
$\left| \lambda _0,\lambda _1,\ldots ,\lambda _N\right\rangle $.
The vacuum states are parametrized by the solutions $\lambda _0$
of the equations (\ref{Aac}). In all cases under consideration the
operator $A_0$ is diagonal while the operator $A$ is weighted
unilateral shift operator. One does not exclude the case when the
irreducible representations generated by $\left| \lambda
_0,\lambda _{1,}\ldots ,\lambda _N\right\rangle $ are of finite
dimension.

We can give now the following
\begin{proposition}If the structural function $\cal G$ is regular
then:
\begin{enumerate}
\item[1)]$\dim {\cal H}
_{\lambda _1\ldots \lambda _N}<\infty $ if  and only if there
exists a pair $i,j\in \left\{0,1,\ldots,N\right\}$ such that
$\frac{k_i}{k_j}<0$.
\item[2)]if $k_0>0$ then the equation  $ A\left| \lambda _0,%
\lambda _1,\ldots ,\lambda _N\right\rangle =0$ is solved by
\begin{equation}
\lambda _{0,l}:=\frac l{k_0}-\frac 1{k_0}{\sum_{j=1}^N} \beta
_{0j}\lambda _j
\end{equation}
where $\beta_{ij}$ are matrix elements of $\alpha^{-1},\;l\in
L:=\left\{0,\frac{1}{\kappa}k_0, \frac{2}{ \kappa} k_0, \ldots
,\frac{\kappa-1}{\kappa}k_0\right\}$ and $\kappa$ is the biggest
common divisor of the numbers ($k_0,k_1, \ldots ,k_N$). Moreover
${\cal H}_{\lambda _1\ldots \lambda _N}$ splits onto the
irreducible components
\begin{equation}
{\cal H}_{\lambda _1\ldots \lambda _N}=\bigoplus_{l\in L}{\cal
H}_{\lambda _1\ldots \lambda _N}^l  \label{Aacc}
\end{equation}
where ${\cal H}_{\lambda _1,\ldots ,\lambda _N}^l$ are generated
by
 ${\cal A} _{red}$ out of the states $\left| \lambda
_{0,l},\lambda _1,\ldots ,\lambda _N\right\rangle $.
\item[3)] $\dim {\cal H}
_{\lambda _1,\ldots ,\lambda _N}=\infty $ if and only if $\dim
{\cal H}^l _{\lambda _1, \ldots, \lambda _N}=\infty $ for all $l$.
\end{enumerate}
\end{proposition}

{\it Proof} : follows immediately from the observations above.
Irreducibility of ${\cal H}^l_{\lambda_1\ldots\lambda_N}$ is a
consequence of the fact that $\left|\lambda_{0,l}, \lambda_1,
\ldots,\lambda_N\right\rangle$ is the unique vacuum in this space.
QED

The construction presented above generalizes the one of
\cite{Kar1},\cite{Kar2}, where the models with ${\cal G}$ being a
polynomial of special type are considered. In these papers the
approximative methods of integration of the models were used.

As it was mentioned in the Introduction, our aim is to study some
integrable family of Hamiltonians (\ref{Azz}). Therefore, instead
of the operator $A_0$, we will use the Hermitian operator
\begin{equation}
Q:=q^{A_0-\lambda _{0,l}} ,  \label{Aad}
\end{equation}
where $0<q<1$.

The structural relations (\ref{Av}-\ref{Ay}) acquire the following
form in terms of the operators $A,A^{*}$ and $Q$
\begin{eqnarray}
QA^{*} &=&qA^{*}Q,\quad qQA=AQ  \label{Aadd} \\ A^{*}A &=&{\cal
R}\left( Q\right)  \label{Aae} \\ AA^{*} &=&{\cal R}\left(
qQ\right) ,  \label{Aaee}
\end{eqnarray}
where structural function ${\cal R}$ is given by
\begin{equation}
{\cal R} \left( Q \right) = {\cal G} \left( \frac{\log Q}{\log q}
+ \lambda_{0,l} -1,\lambda_1, \ldots,\lambda_N \right).
\end{equation}
The function ${\cal R}$ takes positive values in all points
$\left\{ q^n \right\} _{n=1}^{\infty} $ and moreover ${\cal
R}\left( 1 \right) =0$. The algebras of the type above were
analyzed in \cite{Odz}. The reduced Hamiltonian (\ref{Azz}) can be
rewritten as
\begin{equation}
H_{red}={\cal D}\left( Q\right) +A+A^{*},  \label{Aaf}
\end{equation}
with the function ${\cal D}$ given by
\begin{equation}
{\cal D} \left( Q \right) = H_0 \left( \frac{\log Q}{\log q} +
\lambda_{0,l} ,\lambda_1, \ldots,\lambda_N \right).
\end{equation}
by the use
of (\ref{Aad}).

The analysis below is restricted to infinite-dimensional case
only. The discussion of finite dimensional case will be presented
in the separate paper.

\section{Spectral and holomorphic representations}
\renewcommand{\theequation}{2.\arabic{equation}}
\setcounter{equation}{0}

Let $\cal A_R$ be the operator algebra generated by the operators
$A$, $A^{*}$ and $Q$. In this section we describe two natural and,
important from physical point of view, representations of the
algebra ${\cal A}_{{\cal R}}$.

The first representation, which we will call a holomorphic one, is
related to the coherent states $\left| z\right\rangle ,\;\;\left|
z\right| <{\cal R}\left(0\right)$, of the anihilation operator $
A\in {\cal A}_{\cal R}$
\begin{equation}
A\left| z \right\rangle =z\left| z\right\rangle .  \label{Ba}
\end{equation}

Let us consider the case when the orthonormal basis of the Hilbert
space ${\cal H}_{red} := {\cal H}^l_{\lambda _1\ldots \lambda _N}$
\begin{equation}
\left| n\right\rangle :=\left| \lambda _{0,l}+n,\lambda _1,\ldots
,\lambda _N\right\rangle ,\quad \quad \quad \quad \quad n\in {\Bbb
{N}}\cup \left\{ 0\right\}   \label{Bb}
\end{equation}
is infinite. The basis vectors are generated by the operator
$A^{*}$ out of the vacuum state $\left|0\right\rangle$. The action
of the algebra generators on the vectors of this basis is:
\begin{eqnarray}
Q\left| n\right\rangle  &=&q^n\left| n\right\rangle   \label{Bc}
\\ A\left| n\right\rangle  &=&\sqrt{{\cal R}\left( q^n\right)
}\left| n-1\right\rangle   \label{Bd} \\ A^{*}\left|
n\right\rangle  &=&\sqrt{{\cal R}\left( q^{n+1}\right) }\left|
n+1\right\rangle .  \label{Be}
\end{eqnarray}
The coherent state $ \left| z\right\rangle $ is thus given by
\begin{equation}
\left| z\right\rangle :=\sum_{n=0}^\infty \frac{z^n}{\sqrt{{\cal
R}\left( q\right) \ldots {\cal R}\left( q^n\right) }}\left|
n\right\rangle, \label{Bf}
\end{equation}
where $z\in {\Bbb D}=\left\{ z\in {\Bbb C}:\left| z\right| <{\cal
R}\left( 0\right) \right\} $. The states $\left| z\right\rangle $,
$z\in {\Bbb D}$ form a linearly dense subset in ${\cal H}_{red}$.
Therefore, the map
\begin{equation}
I \left( v \right) \stackrel{def}{=} \left\langle v \right. \left|
z \right\rangle \label{Bg}
\end{equation}
where $v\in {\cal H}_{red}$ and $z\in {\Bbb {D}}$ , is an
antilinear and one-to-one map of the Hilbert space ${\cal
H}_{red}$ into the vector space ${\cal O}\left( {\Bbb D}\right) $
of holomorphic functions on the disc ${\Bbb {D}}$ . It was shown
in \cite{Odz} that the image $I\left( {\cal H}_{red}\right) $ is
isomorphic to the Hilbert space\ $L^2{\cal O}\left( {\Bbb
D},d\mu_{\cal R}\right)$ containing holomorphic functions on $
{\Bbb {D}}$ which are square integrable with respect to the
measure
\begin{equation}
d \mu _{\cal R} \left( z, \overline z \right) = \frac{1}{2 \pi}
\frac{1}{Exp_{\cal R} \left(  x \right)} \frac{1}{ \left( 1-q
\right) \left(  q;q \right)_{\infty}} \gamma \left( \frac{1}{x}
\right) \lim_{a \rightarrow{\infty}} \frac{a^{-\frac{\log x}{\log
q}} \left( a;q \right)_{\infty}}{\left( ax;q \right)_{\infty}} d_q
x d\varphi, \label{Bh}
\end{equation}
where $z=\sqrt{x}e^{i\varphi},\;\;d_q x$ is the Jackson measure
(see Appendix A), $d\varphi$ is the Lebesque measure on the circle
$S^1$. The $\cal R$-exponential function $Exp_{\cal R}$ (see
\cite{Odz}) and the function $\gamma$ are defined by
\begin{equation}
\label{Bha}Exp_{\cal R}\left( x \right) := \left\langle z|z \right
\rangle = \sum_{n=0}^{\infty} \frac{x^n}{{\cal R}\left( q
\right)\ldots{\cal R}\left( q^n \right)},
\end{equation}
\begin{equation} \label{Bhh} \gamma\left( z \right) :=
\sum_{n=0}^{\infty} \left( \sum _{k=0}^n \frac{{\cal R}\left(q;q
\right)_{n-k}}{\left( q;q \right)_k} \left( -1 \right)^k q^{\left(
^k_2 \right)}\right)z^k,
\end{equation}
with
\begin{equation}
{\cal R}( q;q )_k := {\cal R}( q ){\cal R}( q^2 )\ldots {\cal
R}(q^{k} ),
 \end{equation}
\begin{equation}
 ( q;q )_k := (1-q )( 1-q^2 )\ldots ( 1-q^{k} ),
\end{equation}
\begin{equation}\left( x;q \right)_{\infty} :=
\prod^{\infty}_{k=0} \left( 1-q^k x \right).\end{equation}
 Let us remark here, that the existence of the morphism (\ref{Bg}) is equivalent to the existence of the resolution of the
unity of the type:
\begin{equation}
\int _{\Bbb D} \left| z\right\rangle \left\langle z\right| d\mu
_{{\cal R}}\left( z, \overline{z}\right) =1.  \label{Bi}
\end{equation}

By holomorphic representation of the algebra ${\cal A}_{{\cal R}}$
we will understand the representation in the Hilbert space
$L^2{\cal O}\left( {\Bbb {D}},d\mu _{\cal R}\right) $.
Straightforward calculation shows that:
\begin{eqnarray}
A\varphi \left( z\right)  &=&\partial _{{\cal R}}\varphi \left(
z\right) \label{Bj} \\ A^{*}\varphi \left( z \right)  &=&z\varphi
\left( z\right)   \label{Bk} \\ Q\varphi \left( z \right)
&=&\varphi \left( qz\right)   \label{Bl} \\ H_{red}\varphi \left(
z\right)  &=&\left( {\cal D}\left( Q\right) +z+\partial _{{\cal
R}}\right) \varphi \left( z\right),   \label{Bm}
\end{eqnarray}
where $\varphi \in L^2{\cal O} \left( {\Bbb {D}},d\mu _{\cal R}%
\right) $ and $\partial _{\cal R}$ is a ${\cal R}$-difference
operator given by
\begin{eqnarray}
\partial _{\cal R}\varphi \left( z \right)  &:=&{\cal R}\left(
q Q \right) \partial _0 \varphi \left( z \right) ,  \label{Bn} \\
\partial _0 \varphi \left( z \right)  &:=& \frac{ \varphi \left(
z \right) -\varphi \left( 0\right) }{z},  \label{Bo}
\end{eqnarray}
see \cite{Odz}. In the cases with
\begin{equation}
{\cal R}\left( x \right) =\frac{1-x}{1-q}  \label{Bp}
\end{equation}
$\partial _{\cal R}$\ is a $q$-derivative $\partial _q$, and the
standard derivative $\frac{d}{d z}$ is obtained from $\partial _q$
in the limit $ q \rightarrow 1$. The quantum algebra ${\cal
A}_{\cal R}$ of ${\cal R}$ given by (\ref{Bp}) is a $q$
-deformation of the Heisenberg algebra. Hence the analytic
realization of  $ {\cal A}_{\cal R}$ introduced above is a natural
generalization of the Bergman-Fock-Segal representation of the
Heisenberg algebra.

The second representation of ${\cal A}_{{\cal R}}$ is related to
the spectral measure of a selfadjoint extension of the Hamiltonian
(\ref{Aaf}). The action of $H_{red}$ on the elements of the
orthonormal basis $\left\{ \left| n\right\rangle \right\}
_{n=0}^{\infty} $ is given in terms of three-diagonal (Jacobi)
matrix
\begin{equation}
H_{red}\left| n\right\rangle =\sqrt{{\cal R}\left( q^n\right)
}\left|
n-1\right\rangle +{\cal D}\left( q^n\right) \left| n\right\rangle +\sqrt{%
{\cal R}\left( q^{n+1}\right) }\left| n+1\right\rangle .
\label{Br}
\end{equation}
We will call this matrix the Jacobi matrix of the operator
$H_{red}$.

The operator $H_{red}$ is symmetric and its domain $D_{H_{red}}$
contains all finite linear combinations of the basis elements
$\left\{ \left| n\right\rangle \right\} _{n=0}^\infty $. The
theory of such type operators is strictly related with the theory
of orthogonal polynomials \cite{A-G},\cite{A},\cite{Ch},\cite{Su}.

Let ${\cal K}_{\omega}$ denote the deficiency subspace of
$H_{red}$ for $\omega \in {\Bbb C}$ and $Im\,\omega \neq 0$
\begin{equation}
{\cal K}_{\omega}:=\left(\left( H_{red}-\omega 1 \right)
D_{H_{red}} \right) ^\bot .\label{Bs}
\end{equation}
The deficiency indices $\left( n_{+},n_{-}\right) $:
\begin{eqnarray}
n_{+} &=&\dim {\cal K}_{\omega} ,\quad for\;\;Im\,{\omega} >0
\label{Bt} \\ n_{-} &=&\dim {\cal K}_{\omega} ,\quad
for\;\;Im\,{\omega} <0 \label{Bu}
\end{eqnarray}
of the operator $H_{red}$ are $\left( 0,0\right) $ or $\left(
1,1\right) $. In order to show this property, one should observe
that $|v\rangle\in {\cal K}_{\overline{{\omega}}}$ if and only if
\begin{equation}
H^{*}_{red}\left| v\right\rangle ={\omega}\left| v\right\rangle.
\label{Bv}
\end{equation}
where $H^*_{red}$ is the Hermitian conjugate of $H_{red}$. The
vector
\begin{equation}
\left| v\right\rangle =\sum_{n=0}^\infty P_n\left( {\omega}
\right) \left| n\right\rangle\in {\cal H}_{red},  \label{Bx}
\end{equation}
solves (\ref{Bv}) if and only if coefficients $ P_n\left({\omega}
\right)$ do satisfy the three term recurrence equation
\begin{equation}
{\omega} P_n\left( {\omega} \right) =\sqrt{{\cal R}\left( q^n\right) }%
P_{n-1}\left( {\omega} \right) +{\cal D}\left( q^n\right)
P_n\left( {\omega} \right) +\sqrt{{\cal R}\left( q^{n+1}\right)
}P_{n+1}\left( {\omega} \right) \label{By}
\end{equation}
$n\in {\Bbb N}$, with the initial conditions
\begin{equation}
P_0\left( {\omega} \right) \equiv 1,\quad \quad P_1\left( {\omega}
\right) =\frac{ {\omega} -{\cal D}\left( 1\right) }{{\cal R}\left(
q\right) } \label{Bz}
\end{equation}
and
\begin{equation}
\sum_{n=0}^\infty \left| P_n\left( {\omega} \right) \right|
^2<+\infty  \label{Baa}
\end{equation}
Hence $n_{+}$ and $n_{-}$ are equal to $0$ or $1$.

Because every $P_n \left(\omega\right)$ is a real polynomial of
degree $n$ of the complex variable ${\omega}$ one has $
n_{+}=n_{-}$.

Following \cite{A} we will call the Jacobi matrix of $H_{red}$ to
be of the type $D$ or $C$ if the deficiency indices of $H_{red}$
are $\left(0,0\right)$ or $\left(1,1\right)$ respectively.

\begin {proposition}
\begin{enumerate}
\item[i)]  If
\begin{equation}
\sum_{n=0}^\infty \frac 1{\sqrt{{\cal R}\left( q^n\right)
}}=+\infty \label{Bae}
\end{equation}
then the operator $H_{red}$ has deficiency indices
$\left(0,0\right)$.This is equivalent to its essential
selfadjointnes.
\item[ii)]  If the set of the coherent states $\left| z\right\rangle $ of
the annihilation operator $A$ is parametrized by the disc ${\Bbb
D}$ of finite radius ${\cal R}\left( 0\right) <+\infty $ then
$H_{red}$ is essentially selfadjoint.

\item[iii)]If the deficiency indices of $H_{red}$ are $\left(
1,1\right) $ then coherent states $\left| z\right\rangle $ of $A$
exist for any $z\in {\Bbb C}$.
\end{enumerate}
\end {proposition}
{\it Proof\,}:
\begin{enumerate}
\item[i)]
Let $\left\{ Q_n\left( {\omega} \right) \right\} _{n=0}^\infty $
be an another solution of the recurrence (\ref{By}) with the
initial conditions given by :
\begin{equation}
Q_0\left( \omega \right) \equiv 0,\quad Q_1\left( \omega \right)
\equiv  \frac 1{\sqrt{ {\cal R}\left( q\right) }}.  \label{Bab}
\end{equation}
One then has
\begin{equation}
P_{k-1}\left( \omega \right) Q_k\left(\omega \right) -P_k\left(
\omega \right) Q_{k-1}\left( \omega \right) =\frac 1{\sqrt{{\cal
R}\left( q^k\right) }}  \label{Bac}
\end{equation}
where $k\in {\Bbb N}$. Applying Schwartz inequality to
(\ref{Bac}), one finds
\begin{equation}
\sum_{n=1}^\infty \frac 1{\sqrt{{\cal R}\left( q^n\right) }}\leq
2\left( \sum_{n=0}^\infty \left| P_n\left( \omega \right) \right|
^2\right) \left( \sum_{n=0}^\infty \left| Q_n\left( \omega \right)
\right| ^2\right) . \label{Bad}
\end{equation}
Moreover one can prove (see \cite{A}) that if $\left\{ P_n\left(
\omega \right) \right\} _{n=0}^\infty $ satisfies (\ref{Baa}),
then $\left\{ Q_n\left( \omega \right) \right\} _{n=0}^\infty $
satisfies it too. Therefore $\sum_{n=0}^\infty \frac 1{\sqrt{{\cal
R}\left( q^n\right) }}<+\infty$. This contradicts (\ref{Bae}).
Thus, one has
\begin{equation}
\sum_{n=0}^\infty \left| P_n\left( \omega \right) \right|
^2=\infty, \label{Baf}
\end{equation}
 meaning that $n_{+}=n_{-}=0$. We apply  Th.VIII.3 from
\cite{R-S} vol.1.

\item[ii)]  If ${\cal R}\left( 0\right) <+\infty $ then
the condition  (\ref{Baf}) follows immediately from (\ref{Bae}).
The statement ii) follows from the statement  i).

\item[iii)]One proves it {\it ad absurdum} using the previous statement. QED
\end{enumerate}

Let $\widehat{H}_{red}$ be some selfadjoint extension of
$H_{red}$. In the case of type $C$ such extensions are
parametrized by points of $S^1$ see \cite{R-S} whereas in the case
of type $D$ the extension is unique. Let $dE_{\cal
R,D}\left(\omega\right)$ be the spectral measure of
$\widehat{H}_{red}$.

By the spectral representation of $\cal A_R$ we will call the
representation in the space $L^2\left({\Bbb R},d\sigma_{\cal
R,D}\right)$ where the measure $d\sigma_{\cal R,D}$ is given by
\begin{equation}
d\sigma_{\cal R,D}\left(\omega\right):=\left\langle0|dE_{\cal
R,D}\left(\omega\right)0\right\rangle
\end{equation}
and the operator $\widehat{H}_{red}$ acts by multiplication with
the identity function on $\Bbb R$. Let $U_{PN}:{\cal
H}_{red}\longrightarrow L^2\left({\Bbb R},d\sigma_{\cal
R,D}\right)$ denote the intertwining operator for these two
representations. Because the polynomials $\left\{ P_n \left(
\omega \right) \right\}_{n=0}^\infty$ form an orthonormal basis in
$L^2\left( {\Bbb R},d\sigma_{\cal R,D} \right)$, it is convenient
to write the intertwining operator using the physical notation of
Dirac
\begin{equation}
\label{Bah} U_{PN}=\sum^\infty_{n=0}P_n \left( \omega \right)
\otimes\left\langle n \right|.
\end{equation}
The convergence is understood in the sense of weak topology. The
relation of the holomorphic representation and the spectral
representation is given by the isomorphism of Hilbert spaces
\begin{equation}
U_{PZ}:L^2 {\cal O} \left( {\Bbb D},d\mu_{\cal R} \right)
\rightarrow L^2 \left( {\Bbb R},d\sigma_{\cal R,D} \right)
\end{equation}
and $U_{PZ}$ one can be written as
\begin{equation}
U_{PZ}=\sum^\infty_{n=0} P_n\left( \omega  \right)\otimes
\frac{z^n}{\sqrt{{\cal R}\left(q\right) \ldots {\cal R}\left(q^n
\right)}},
\end{equation}
where $\left\{\frac{z^n}{\sqrt{{\cal R}\left(q\right)\ldots {\cal
R}\left(q^n \right)}}\right\}_{n=0}^\infty$ is an orthonormal
basis in $L^2 {\cal O} \left( {\Bbb D},d\mu_{\cal R} \right)$.
Similarly let
\begin{equation}
U_{ZN}:{\cal H}_{red}\longrightarrow
L^2 \left( {\Bbb D},d\mu_{\cal R} \right)
\end{equation}
be given by
\begin{equation}
U_{ZN}:=\sum^\infty_{n=0} \frac{z^n}{\sqrt{{\cal R}\left(q\right)
\ldots {\cal R}\left(q^n \right)}} \otimes \left\langle n\right|.
\end{equation}

We are thus getting the following commutative diagram
\begin{eqnarray}
&{\cal H}_{red}& \nonumber \\ \stackrel{U_{PZ}}{\swarrow} &&
\stackrel{U_{ZN}}{\searrow}\\ L^2\left( {\Bbb R}, d\sigma_{\cal
R,D} \right)&\stackrel{U_{PN}}{\longleftarrow}&  L^2 {\cal O}
\left( {\Bbb D}, d\mu_{\cal R} \right)\nonumber
\end{eqnarray}
of Hilbert space isomrphisms.

If the series
\begin{equation}
V\left(\omega,z\right):=\sum^\infty_{n=0} P_n\left( \omega \right)
\frac{z^n}{\sqrt{{\cal R}\left(q\right) \ldots {\cal R}\left(q^n
\right)}}
\end{equation}
is pointwise convergent for all $\omega\in\left[a,b\right]$ and
$z\in \Bbb D$, the isomorphism $U_{PZ}$ can be represented as the
integral transform
\begin{equation}
\label{Bai}
 \left(U_{PZ}\varphi \right)\left( \omega
 \right)=\int_{\Bbb D}
 V\left(\omega,z\right)\varphi\left(z\right)d\mu_{\cal
 R}\left(z,\overline{z}\right).
\end{equation}
The kernel $V\left(\omega,z\right)$ of this transform satisfies
the ${\cal R}$-difference equation
\begin{equation}
\label{Baj} \left( \omega-z \right)V \left( \omega,z \right)={\cal
D } \left( Q \right)V \left( \omega,z \right)+\partial_{\cal R}V
\left( \omega,z \right).
\end{equation}
Of course, from the orthogonal polynomials theory point of view $V
\left( \omega,z \right)$ is nothing else than the generating
function for the family of orthogonal polynomials under
consideration.

The function $V\left(\omega,z\right)$ is useful to calculate many
important physical quantities of the system described by the
Hamiltonian $H_{red}$.\\
 First of all, note that using (\ref{Bai})
and (\ref{Baj}) one obtains
\begin{equation}
\left\langle v\right.\left|H_{red}^n z\right\rangle =\int_{\Bbb
R}\overline{V \left( \omega,v \right)}\omega^n V \left( \omega,z
\right)d\sigma_{\cal R,D}\left( \omega \right)= \left( z+{\cal
D}\left( Q \right)+\partial_{\cal R} \right)^n Exp_{\cal R}\left(
\overline{v}z \right).
\end{equation}
Putting in particular  $n=0$ we have
\begin{equation}
Exp_{\cal R}\left( \overline{v}z \right)=\int_{\Bbb R}\overline{V
\left( \omega,v \right)} V \left( \omega,z \right)d\sigma_{\cal
R,D}\left( \omega \right)
\end{equation}
which is the integral representation of the $\cal R$-exponential
function $Exp_{\cal R}\left( \overline{v}z \right)$. This function
satisfies the equation
\begin{equation}
\partial_{\cal R} Exp_{\cal R} \left( \overline{v} z \right)%
= \overline{v}Exp_{\cal R}\left( \overline{v}z \right).
\end{equation}
Let us recall that $Exp_{\cal R}\left(\overline{v}\cdot\right)\in
L^2{\cal O}\left({\Bbb D},d\mu_{\cal R }\right)$ is the expression
of the coherent state in holomorphic representation.

The evolution operator $U\left(t\right):=e^{i\widehat{H}_{red}t}$
acts on the function from $L^2\left({\Bbb R},d\sigma_{\cal R,D
}\right)$ as multiplication  by phase factors:
\begin{equation}
U\left(t\right) \psi\left(\omega\right)=e^{i\omega t}
\psi\left(\omega\right).
\end{equation}
It enables us to calculate the transition amplitudes between
coherent states :
\begin{equation}
\left\langle v|U\left(t\right)z\right\rangle=\int_{\Bbb
R}\overline{V\left(\omega,v\right)}e^{i\omega t}
V\left(\omega,v\right)d\sigma_{\cal R,D}\left(\omega\right).
\end{equation}
The vacuum - vacuum transition amplitude is also important for
physicists. It is given by:
\begin{equation}
\left\langle 0|U\left(t\right)0\right\rangle=\int_{\Bbb
R}e^{i\omega t}d\sigma_{\cal
R,D}\left(\omega\right)=\sum^\infty_{n=0}\frac{\left(it\right)^n}{n!}\mu_n
\end{equation}
where $\mu_n$ is the n-th moment of the measure $d\sigma_{\cal
R,D}$.

The above shows that the measure $d\sigma_{\cal R,D}$ plays a
significant role in the description of our physical system. The
construction $d\sigma_{\cal R,D}$ is one of the most important
problems which has to be solved in order to recover the dynamics.\newline%
In order to give an example see \cite{A} of the solution of this problem, let
us recall the notion of a simple symmetric operator.

The Hilbert subspace ${\cal H}_1\subset{\cal H}$ is a reducible
subspace of the linear operator $T:{\cal H}\longrightarrow{\cal
H}$ if ${\cal H}_1$ and ${\cal H}_2:={\cal H}_1^\bot$ are
invariant subspaces for $T$ and for orthogonal projection
$\Pi_1:{\cal H}\longrightarrow {\cal H}_1$ one has
$\Pi_1\left(D_T\right)\subset D_T$. The symmetric operator $T$ is
simple if there does not exist the irreducible subspace of $T$
such that $T_{\mid{\cal H}_1}$ has selfadjoint extension in
${{\cal H}_1}$.

If the Jacobi matrix of the reduced Hamiltonian $H_{red}$ is of
type $C$ then the series
\begin{equation}
\sum_{n=0}^\infty\left|P_n\left(\omega\right)\right|^2\;\;\; \;
{\&} \; \; \; \; \sum_{n=0}^\infty \left| Q_n \left( \omega
\right) \right|^2
\end{equation}
are almost uniformly convergent on $\Bbb C$. Thus the functions
$A\left(\omega\right),B\left(\omega\right),C\left(\omega\right)$
and $D\left(\omega\right)$ defined by
\begin{eqnarray}
 A\left(\omega\right) &=& \omega \sum_{k=0}^\infty
Q_k\left(0\right)Q_k\left(\omega \right),\nonumber \\
B\left(\omega\right)&=& -1+ \omega \sum_{k=0}^\infty
Q_k\left(0\right)P_k\left(\omega \right),\nonumber
\\ \label{Baff} C\left(\omega\right)&=& 1+ \omega
\sum_{k=0}^\infty P_k\left(0\right)Q_k\left(\omega \right),\\
D\left(\omega\right)&=& \omega \sum_{k=0}^\infty
P_k\left(0\right)P_k\left(\omega\right)\nonumber
\end{eqnarray}
are entire  functions. If in addition $H_{red}$ is simple and closed operator then the
spectral measure $dE_{\cal R,D}\left(\omega\right)$ of its
arbitrary  selfadjoint extension $\widehat{H}_{red}$ is localized
at the nulls $\omega_i,\;\;i=1,2,\ldots$, of the function
$q\left(\omega\right)=B\left(\omega\right)t-D\left(\omega\right)$
for some $t\in \Bbb R$. The steps $\mu_i,\;i=1,2,\ldots$ of the
measure $d\sigma _{\cal R,D}\left(\omega\right) :=\left\langle
0|dE_{\cal R,D}\left(\omega\right)0 \right\rangle$ satisfy the
following conditions
\begin{eqnarray}
\sum_{i=1}^\infty\frac{1}{\mu_i \left(1+\omega^2_i \right)\left|
q'\left(\omega_i\right) \right|^2}<\infty\\
\sum_{i=1}^\infty\frac{1}{\mu_i \left| q'\left(\omega_i\right)
\right|^2} = \infty.
\end{eqnarray}

From the identity
\begin{equation}
\left|n \right\rangle=P_n\left(H_{red}\right)\left|0 \right\rangle
\end{equation}
it follows that if the Jacobi matrix of $H_{red}$ is of type $D$
then $H_{red}$ has simple spectrum. Conversely, one can prove (see
\cite{St}) that every selfadjoint operator $H=H^*$ with simple
spectrum may be represented in some orthonormal basis by the
formula (\ref{Br}) where the Jacobi matrix is of the type $D$.

Using (\ref{Bc}-\ref{Be}) we can associate with it the algebra
$\cal A_R$ with the proper structural function $\cal R$. This fact
indicates that the algebras of this kind are important tools to
investigate the symmetry structures of the physical systems with
dynamics generated by Hamiltonians with simple spectrum.

In the next section we will describe the situation when neither
coherent states nor the kernel $V\left(\omega,z\right)$ do exist.


\section{The integrable systems related to q-Hahn's polynomials}
\renewcommand{\theequation}{3.\arabic{equation}}
\setcounter{equation}{0}

 We will integrate quantum systems related
to the algebras ${\cal A}_{\cal R}$ and Hamiltonian
$H_{red}=:H_{AB}$ with the corresponding structural functions of
the form
\begin{eqnarray}
 \label{Ca} &{\cal R}_{AB}\left(x\right)= \left(1-q\right)x
 \left[\partial _q\eta\left(x\right)-\beta\left(x\right)\partial _q
 \beta\left(x\right)\right]&\\ \label{Cb}&{\cal
 D}_{AB}\left(x\right)= \left(1-q\right)x
 \partial _q \beta\left(x\right).&
\end{eqnarray}
where
\begin{equation}
 \label{Cc}
 \beta \left(x\right) = \frac{q\left(1-x\right)\left[ \left(a_0
 \left(1-q \right)-b_1 \right)x + b_1
 q\right]}{\left(1-q\right)\left[\left(a_1\left(1-q \right) -b_ 2
 \right)x^2 + b_2q^2 \right]}
 \end{equation}
 \begin{equation}
 \label{Ccc} \eta \left( x \right) = \left\{\frac{\left[\left( a_0
 \left( 1-q \right) -b_1 \right) x + b_1 q^2 \right]{\left[\left(
 a_0 \left( 1-q \right) -b_1 \right) x + b_1 q
 \right]}}{{\left[\left( a_1 \left( 1-q \right) -b_2 \right) x^2 +
 b_2 q^2 \right]}\left[\left( a_1 \left( 1-q \right) -b_2 \right)
 x^2 + b_2 q^3 \right]}+\right.
\end{equation}
$$
 +\left. \frac{b_0 \left( 1-q \right)}{\left( a_1 \left( 1-q
 \right) -b_2 \right) x^2 + b_2 q^3 }\right\}\frac{q^3
 \left(1-x\right)\left(1-q^{-1}x\right)}{\left(1-q\right)^2
 \left(1+q\right)}.
$$
 These functions depend on  five real parameters
$a_0,a_1$ and $b_0,b_1,b_2$. It will appear later on that it is
natural to introduce the following two polynomials:
\begin{eqnarray}
 \label{Cd}A\left(\omega \right) &=&a_1\omega+a_0\\
 \label{Ce}B\left(\omega \right)&=&b_2\omega^2+b_1\omega+b_0.
\end{eqnarray}
It is clear from (\ref{Cc}-\ref{Ccc}) that the pairs of
polynomials
$\left(A\left(\omega\right),B\left(\omega\right)\right)$ of degree
one and two respectively, taken up to common overall real factor
$c\neq0$, parametrize the models under consideration. The only
condition we will impose on the pair
$\left(A\left(\omega\right),B\left(\omega\right)\right)$ is the
one given by
 ${\cal R }_{AB}\left(q^n\right)>0$ for any $n\in{\Bbb N}$.

Analogously to the theory of classical orthogonal polynomials
there is a $q$-difference equation (an analog of Pearson equation
\cite{Su}):
\begin{equation}
 \label{Cf}\partial _q \left( \varrho B \right)\left( \omega
 \right) = \left( \varrho A \right) \left( \omega \right)
\end{equation}
associated with the pair ($A\left(\omega\right), B \left( \omega
\right)$).

We will look for the solutions $\varrho \left( \omega \right)$ of
(\ref{Cf}) which do satisfy the boundary conditions
\begin{equation}
 \label{Cg} \varrho \left( a \right) B \left( a \right) =  \varrho
 \left( b \right) B \left( b \right) = 0,
\end{equation}
for some fixed $a,b$ such that $-\infty \leq a<b\leq\infty$. We
thus have so called Pearson data
$\left(A\left(\omega\right),B\left(\omega\right)\right)$ on the
interval $(a,b)\subset{\Bbb R}$.
\begin{proposition}
Let $\varrho \left( \omega \right)$ be the solution of the
$q$-Pearson equation (\ref {Cf}) defined by
$\left(A\left(\omega\right), B \left( \omega \right)\right)$ and
satisfying (\ref{Cg}). Then $\varrho ^{\left(k\right)} \left(
\omega \right) :=\varrho \left(q^k \omega
\right)B\left(q\omega\right)\ldots B\left(q^k\omega\right)$ is the
solution of Pearson $q$-equation (\ref{Cf}) associated with the pair
 \begin{eqnarray}
 \label{Ch} A^{(k)}(\omega) &:=& q^k A
 (q^k \omega)+\frac{1-q^kQ^k}{1-q}\partial _q B(\omega)\\ \label{Ci} B^{\left(k\right)}   \left(\omega\right)
 &:=& B \left(\omega\right).
 \end{eqnarray}
where $k\in {\Bbb N}$. If $\varrho(\omega)$ satysfies the boundary conditions (\ref{Cg}) then $\varrho^{(k)}(\omega)$ satysfies (\ref{Cg}) too.
\end{proposition}
{\it Proof} : By straight forward calculation.  QED

Let $L^2 \left( \left[a,b\right] ,\;d \sigma_{AB} \right)$ be the
Hilbert space of square-integrable functions with respect to the
measure
 \begin{equation}
 \label{Cj}d\sigma_{AB} \left( \omega \right) = \varrho \left(
 \omega \right) d_q \omega
 \end{equation}
where
 \begin{equation}
 \label{Ck}d_q \omega  = \sum ^\infty _{k=0} \left( 1-q \right)q^k
 \left[ b \delta \left(\omega-q^k b \right) - a \delta
 \left(\omega-q^k a \right) \right] d \omega
 \end{equation}
is the Jackson measure on the interval $\left[ a,b \right]$. It
will be assumed that the weight function $\varrho \left(
\omega\right)$ does satisfy the Pearson $q$-equation (\ref{Cf})
supplemented with boundary condition (\ref{Cg}).

Applying the orthonormalization procedure to the monomials
$\left\{\omega^n \right\}^\infty_{n=0}\subset L^2 \left(
\left[a,b\right] ,\;d \sigma_{AB} \right)$  we obtain the system
of orthonormal polynomials (OPS) $\left\{P_n \left( \omega \right)
\right\} ^{\infty} _{n=0}$
 \begin{equation}
 \label{Cl} \int_a^b P_n \left(\omega \right) P_m \left(\omega
 \right) d\sigma_{AB}\left(\omega\right) = \delta_{nm}
 \end{equation}
which is uniquely determined  by the pair  $\left(A \left(\omega
\right),B \left(\omega \right) \right)$ and interval. The
polynomials $\left\{P_n \left( \omega \right) \right\} ^{\infty}
_{n=0}$ are called q-Hahn's polynomials, see \cite{G-R}, \cite{H}.
Let us denote by $\left\{\widetilde{P_n} \left( \omega \right)
\right\} ^{\infty} _{n=0}$ the monic OPS associated with
$\left\{P_n \left( \omega \right) \right\} ^{\infty} _{n=0}$.
(e.i. $\widetilde{P}_n\left(\omega\right)=\frac{1}{\alpha_n}P
_n\left(\omega\right)$ where $\alpha_n$ is the coefficient of the
higest power in $P_n\left(\omega\right)$).
 \begin{theorem}
 If $\left\{\widetilde{P_n} \left( \omega \right) \right\}
 ^{\infty} _{n=0}$ is the monic OPS corresponding to the  Pearson
 data $\left( A \left( \omega \right), B \left( \omega
 \right)\right)$ then the family of polynomials:
  \begin{equation}
  \label{Cm} \left\{\frac{1}{\left[n \right]\left[n-1
  \right]\ldots\left[n-k \right]}\partial_q ^k \widetilde{P_n}
  \left( \omega \right) \right\} ^{\infty} _{n=0}
  \end{equation}
 where $$ \left[k\right]:=\frac{1-q^k}{1-q},$$ forms the monic OPS
 corresponding to $\left( A^{\left( k \right) }\left( \omega
 \right), B^{\left( k \right) } \left( \omega \right)\right)$, with
 the same boundary conditions (\ref{Cg}).
 \end{theorem}
{\it Proof} : For  $k\leq n-2$ we have
 \begin{equation}
 \int _a^b \widetilde{P}_n\left( \omega\right) \omega^kA\left(
 \omega\right) \varrho \left( \omega\right) d_q\omega=0.
 \end{equation}
Using Leibnitz rule, (\ref{Cf}) and (\ref{Cg}) we obtain
 \begin{eqnarray*}
 0&=&\int_a^b \widetilde{P}_n\left( \omega\right) \omega^k\left(
 \partial _qB\varrho \right) \left( \omega\right) d_q\omega=\\
 &=&\left. \widetilde{P}_n\left( \omega\right) \omega^kB\left(
 \omega\right) \varrho \left( \omega\right) \right|
 _a^b-\int_a^b\partial _q\left( \widetilde{P}_n\left( \omega\right)
 \omega^k\right) B\left( q\omega\right) \varrho \left(
 q\omega\right) d_q\omega= \\ &=&-\int_a^b\partial _q\left(
 \widetilde{P}_n\left( \omega\right) \omega^k\right) \left( B\left(
 \omega\right) -\left( 1-q\right) \omega A\left( \omega\right)
 \right) \varrho \left( \omega\right) d_q\omega=\\ &=&-\int_a^b
 \partial _q\widetilde{P}_n \left( \omega\right) \left( q\omega\right) ^k\left(
 B\left( \omega\right) -\left( 1-q\right)\omega A\left(
 \omega\right) \right) \varrho \left( \omega\right) d_q\omega-
 \\ &-&\int_a^b \widetilde{P}_n\left( \omega\right) \left[
 k\right] \omega^{k-1}\left( B\left( \omega\right) -\left(
 1-q\right) \omega A\left( \omega\right) \right) \varrho \left(
 \omega\right) d_q\omega.
 \end{eqnarray*}
The degree of polynomial
 \begin{equation}
 \left[ k\right] \omega^{k-1}\left( B\left( \omega\right) -\left(
 1-q\right) \omega A\left( \omega\right) \right)
 \end{equation}
is $k+1<n$, and from (\ref{Ch}-\ref{Ci}) one finds that:
 \begin{equation}
 \int_a^b \partial _q\widetilde{P}_n\left( \omega\right)
 \omega^k\varrho ^{\left(1\right)}\left( \omega\right) d_q\omega=0
 \end{equation}
for $k\leq n-2$. This shows that the polynomials $\left\{ \frac
1{\left[ n\right] }\partial _q\widetilde{P}_n\left( \omega\right)
\right\} _{n=1}^\infty $ form a monic OPS for the Pearson data
$\left( A^{\left( 1\right) }\left( \omega\right) ,B^{\left(
1\right) }\left( \omega\right) \right) $ given by
(\ref{Ch}-\ref{Ci}).  QED

Since, for the rational function ${\cal R}_{AB}\left( x \right)$
of (\ref{Ca}) one has ${\cal R}_{AB} \left( 0 \right) = 0$  the
{\bf Proposition 2.1} and it's consequences described in the
previous section imply that in the case under
 consideration the following statements are true:
 \begin{enumerate}
 \item[i)] the operator $\overline{H}_{AB}$ is selfadjoint and has
 simple spectrum;
 \item[ii)] the coherent states do not exist for ${\cal A}_{{\cal R}_{AB}}$
 \item[iii)] the Hilbert space ${\cal H}_{red}$ is unitary isomorphic to
 $L^2 \left( \left[ a,b \right] ,\; d\sigma_{AB} \right)$, with the
 isomorphism given by (\ref{Bah}).
 \end{enumerate}
Hence, the Hahn's polynomials $\left\{P_n ( \omega ) \right\}
^{\infty} _{n=0}$ form an orthonormal basis in $L^2 \left( \left[
a,b \right] ,\; d\sigma_{AB} \right)$. The measure $ d\sigma_{AB}$
is the expectation value of the spectral measure $dE_{AB}$ in the
vacuum state $\left| 0 \right\rangle$.

It is then clear that the properties of Hahn's polynomials are
crucial for better understanding  of the physical systems
corresponding the algebra ${\cal A}_{{\cal R}_{AB}}$. The
proposition below gives a description of some important properties
of these polynomials, namely
 \begin{theorem}[Hahn]\label{Hahn}
 Fix a Pearson data $(A ( \omega ),B ( \omega ) )$ on the interval
 $\left[a,b\right]\subset{\Bbb R}$. Then the following statements
 are equivalent:
  \begin{enumerate}
  \item[A.] The family of polynomials $ \left\{ \widetilde{P_n} \left( \omega \right)
  \right\} ^{\infty}
  _{n=0}$ forms the monic OPS with respect to $\left(A\left( \omega
  \right),B \left( \omega \right)\right) $.
  \item[B.] The polynomials are given by formula of Rodriques:
   \begin{equation}
   \label{Cn} \widetilde{P_n} \left( \omega \right)=c_n
   \frac{1}{\varrho \left(\omega \right)}\partial_q^n \left[\varrho
   \left( \omega \right)B\left(\omega\right)B\left(q^{-1}\omega
   \right)\ldots B \left(q^{-\left( n-1 \right)}\omega\right) \right]
   \end{equation}
  $n\in{\Bbb N}$, where $c_n$ is a normalization constant.
  \item[C.] The polynomials $ \left\{ \widetilde{P_n} \left( \omega \right)
  \right\} ^{\infty}_{n=0}$ do satisfy the following $q$-difference
  equation (Hahn equation)
   \begin{equation}
   \label{Co} \left( A \left( \omega \right)\partial_q + B\left(
   \omega \right)
   \partial_q Q^{-1} \partial_q \right) \widetilde{P_n} \left( \omega
   \right)= \lambda_n \widetilde{P_n} \left( \omega \right)
   \end{equation}
  where
   \begin{eqnarray}
   \label{Cp} \lambda_n&=&a_1 \left[ n \right] + b_2 \left[ n \right]
   \left[ n-1 \right]q^{-\left( n-1 \right)};\;\;\;n=2,3,\ldots\\
   \lambda_1&=&a_1.\nonumber
   \end{eqnarray}
  \item[D.] Every polynomial of the system $ \left\{ \widetilde{P_n} \left( \omega \right)
  \right\} ^{\infty}_{n=0}$ is given by
   \begin{equation}
   \label{Cr} \widetilde{P_n} \left( \omega \right) = \prod
   ^{n-1}_{k=0} \frac{1}{a_1 ^{\left( k \right)} - b_2 \left[-n+1+k
   \right]} \left( A^{\left( 0 \right)} \left( \omega \right)+B
   \left( \omega \right)\partial_q Q^{-1} \right) \ldots \left(
   A^{\left( k-1 \right)} \left( \omega \right)+B \left( \omega
   \right)\partial_q Q^{-1} \right) \cdot1
   \end{equation}
 where the linear functions
  \begin{equation}
  \label{Cs}
  A^{ \left( k \right)}\left( \omega \right)=a^{\left( k \right)}_1\omega+a^{\left( k \right)}_0
  \end{equation}
  are defined in (\ref{Ch}).
 \item[E.] The polynomials of the system $ \left\{ \widetilde{P_n} %
 \left( \omega \right)
 \right\} ^{\infty}_{n=0}$ are related by the three term recurrence
 formula
   \begin{equation}
   \label{Ct} \widetilde{P}_{n+1} \left( \omega \right)+ {\cal
   R}_{AB} \left( q^n \right) \widetilde{P}_{n-1} \left( \omega
   \right) = \left( \omega - {\cal D} _{AB}\left( q^n \right) \right)
   \widetilde{P}_n \left( \omega \right)
   \end{equation}
  with the initial condition $P_0\left( \omega \right)\equiv1$.
  \end{enumerate}
 \end{theorem}

The proofs of the equivalence of $A,B,C$ may be found in the
original paper of Hahn \cite{H}. The recurrence formula (\ref{Ct})
is considered there without specification of the form of the
structural functions ${\cal R}_{AB},{\cal D}_{AB}$. The only
assumption made is that they are rational functions of the
parameter $q$. A complete proof  of this theorem is given in the
Appendix B.

From the considerations of Section 2, it follows that the problem
of integration of  the multiboson system described in $L^2\left(
\left[a,b\right] ,d\sigma_{AB}\right)$, is reduced to the
construction of the measure $d\sigma_{AB}$. According to
(\ref{Cj}) the measure $d\sigma_{AB}$ is given by the density
function $\varrho\left(\omega\right)$ which is a solution of the
$q$-difference Pearson equation (\ref{Cf}). Let us therefore
present all possible solutions of (\ref{Cf}) from the class of the
meromorphic functions. Using (\ref{Cd}-\ref{Ce}) we can rewrite
(\ref{Cf}) in the form
 \begin{equation}\label{Cu}
 \varrho(\omega) =\frac{B(q\omega)}{B(\omega)-(1-q)\omega A(\omega)} \varrho(q\omega)=
 \frac{b_2q^2\omega^2+b_1q\omega+b_0}{\left(b_2-\left(1-q\right)a_1\right)
 \omega^2+\left(b_1-\left(1-q\right)a_0\right)\omega+b_0}
 \varrho\left(q\omega\right)
 \end{equation}
and after standard calculations we obtain the following classes of
solutions depending on the values of the parameters
$b_2,b_1,b_0,a_1,a_0$:

\begin{proposition}
One has the following subcases of the solutions of the $q$-difference Pearson %
 equation (\ref{Cf}):

 \begin{enumerate}
  \item[\bf i)]
 If $b_0\neq0$ and $\;b_2-(1-q)a_1\neq0$, then
  \begin{equation}
  \varrho (\omega)
  =\frac{(\frac{q\omega}{a};q)_{\infty}(\frac{q\omega}{b};q)_{\infty}}{(
  \frac{\omega}{c};q)_{\infty}(\frac{\omega}{d};q)_{\infty}},
  \end{equation}
 where $a\neq0,\;b\neq0$ are roots of the polynomial $B(\omega)$ and%
 $\;c\neq0,\;d\neq0$ are roots of the polynomial  $\;B(\omega)-(1-q)\omega A(\omega)$.

 \item [\bf ii)]
 If $b_0\neq0$ and $\;b_1-(1-q)a_0\neq0\;$ and $\;b_2-(1-q)a_1=0\;$, then
  \begin{equation}
  \varrho (\omega)
  =\frac{(\frac{q\omega}{a};q)_{\infty}(\frac{q\omega}{b};q)_{\infty}}{(
  \frac{\omega}{c};q)_{\infty}},
  \end{equation}
 where $a\neq0,\;b\neq0$ are roots of the polynomial $\;B(\omega)\;$ and%
$\;c\neq0\;$ is the root of the polynomial
$\;B(\omega)-(1-q)\omega A(\omega)$.

 \item [\bf iii)]
 If $b_0\neq0$ and $\;b_1-(1-q)a_0=0\;$ and $\;b_2-(1-q)a_1=0\;$, then
  \begin{equation}
  \varrho (\omega)
  =(\frac{q\omega}{a};q)_{\infty}(\frac{q\omega}{b};q)_{\infty},
  \end{equation}
 where $\;a\neq0,\;b\neq0\;$ are roots of the polynomial $\;B(\omega)$.

 \item [\bf iv)]
 If $\;b_0=0,\;\; b_1\neq0\;$ and  $\;b_1-\left(1-q\right)a_0\neq0\;\;$%
  and  $\;\;b_2-\left(1-q\right)a_1\neq0\;$ and $\;b_2\neq0\;$, then
   \begin{equation}
   \varrho (\omega) =\omega^r \frac{(\frac{q\omega}{a};q)_\infty}
   {(\frac{\omega}{c};q)_\infty},
   \end{equation}
where $\;a\neq0\;$ is the root of the polynomial $\;B(\omega)$,
$\;c\neq0\;$ is the root of the polynomial
$\;B(\omega)-(1-q)\omega A(\omega)\;$ and  $q^{-r}= \left|\frac{q
b_1}{b_1-\left(1-q\right)a_0}\right|$.\\

  \item [\bf v)]
   If $\;b_0=0,\;\; b_1\neq0\;$ and  $\;b_1-\left(1-q\right)a_0\neq0\;\;$%
  and  $\;\;b_2-\left(1-q\right)a_1=0\;$ and $\;b_2\neq0\;$, then
   \begin{equation}
   \varrho (\omega) =\omega^r (\frac{q\omega}{a};q)_\infty,
   \end{equation}

where $a\neq0$ is the root of the polynomial $B(\omega)\;$ and
$\;q^{-r}= \left|\frac{q b_1}{b_1-\left(1-q\right)a_0}\right|$.\\

 \item [\bf vi)]
If $b_0=b_1- \left(1-q \right)a_0 =0,\;\;b_1\neq 0,\;\;b_2\neq
0\;$ and $\;b_2- (1-q )a_1 \neq 0$, then
  \item [\bf a)]
   \begin{equation}
   \varrho (\omega)= \omega^r\frac{
   (\frac{q\omega}{a};q)_\infty}{(-\omega;q)_\infty(-q\omega^{-1};q)_\infty}
   \end{equation}
for $q^{-r} =\frac{qb_1}{b_2-\left(1-q\right)a_1}>0$;\\
  \item [\bf b)]
   \begin{equation}
   \varrho (\omega)= \omega^r\frac{
   (\frac{q\omega}{a};q)_\infty}{(\omega;q)_\infty(q\omega^{-1};q)_\infty}
   \end{equation}
for $-q^{-r} =\frac{qb_1}{b_2-\left(1-q\right)a_1}<0$,\\
 where $a\neq0$ is the root of the polynomial $\;B(\omega)$.\\

  \item [\bf vii)] If $b_0=b_1=0,\;\;b_1-(1-q)a_0\neq0$ and
  $b_2\neq0$, then
  \item [\bf a)]
   \begin{equation}
   \varrho (\omega) =\omega^r\frac{(-\omega;q)_\infty(-q\omega^{-1};q)_\infty}{
   (\frac{\omega}{c};q)_\infty}
   \end{equation}\

for $\;q^{-r}= \frac{q^2 b_2}{b_1-\left(1-q\right)a_0}>0$;\\
  \item [\bf b)]
   \begin{equation}
   \varrho (\omega) =\omega^r\frac{(\omega;q)_\infty(q\omega^{-1};q)_\infty}{
   (\frac{\omega}{c};q)_\infty}
   \end{equation}\

for $\;-q^{-r}= \frac{q^2 b_2}{b_1-\left(1-q\right)a_0}<0$,\\
where $c\neq0$ is the root of the polynomial
$\;B(\omega)-(1-q)\omega A(\omega)$.

 \item [\bf viii)]
If $b_0=b_1=b_1- \left(1-q \right)a_0 =0,\;\;$ and $\;b_2- (1-q )a_1 \neq0\;$%
 and  $b_2\neq0$, then
   \begin{equation}
   \varrho (\omega)=\omega^r
   \end{equation}

for
$\;q^{-r}=\left|\frac{q^2b_2}{b_2-\left(1-q\right)a_1}\right|$.\\

  \end{enumerate}
 \end{proposition}

{\it Proof} : The subcases i),  ii) and iii) are easily obtained
by iteration. The points iv)- viii) are proved by calculation of
Laurent expansion coefficient and application of the Ramanujan's
identities (see \cite{G-R}).  QED\\

We can now determine the interval of integration in (\ref{Cl}) and
determine the conditions on polynomials $A(\omega)\;$ and
$\;B(\omega)\;$ such that the measure $d\sigma_{AB}$ is positive
(ie $R(q^n)>0$ for $n \in \Bbb N$). It will be convenient to
express the conditions on $A(\omega)\;$ and $\;B(\omega)\;$ in
terms of roots of the polynomials $\;B(\omega)\;$ and
$B(\omega)-(1-q)\omega A(\omega)$.

\begin{proposition}
 The measure  $d\sigma_{AB}$ is positive and the condition (\ref{Cg}) is
  fulfilled if and only if (in the notation and classification of {\bf Proposition 3.2})
\begin{enumerate}
  \item [{\bf i)}] The integration interval is $[a,b]$ with
  $\;a<0<b\;$ and $\;c,\;d$ satisfies one of the following conditions:
    \subitem {\bf $\alpha$)} $c=\overline{d}$,
    \subitem {\bf $\beta$)} $c<a$ and $d>b$,
    \subitem {\bf $\gamma$)} $c,d<a$,
    \subitem {\bf $\delta$)}  there exists $K\in \Bbb N$ such that %
    $q^{K-1}a<c,d<q^K a$,
    \subitem {\bf $\epsilon$)} $c,d>b$,
    \subitem {\bf $\zeta$)}  there exists $K\in \Bbb N$ such that %
    $q^K b<c,d<q^{K-1}b$.
\item [{\bf ii)}] The integration interval is $[a,b]$ with
$a<0<b$ and  $c<a$ or $c>b$.
\item [{\bf iii)}] The integration interval is $[a,b]$ with
$a<0<b$.
\item [{\bf iv)}] This case splits into two subcases:
    \subitem {\bf 1)} For $a>0$  the integration interval is $[0,a]$
    and $c<0$ or $c>a$.
    \subitem {\bf 2)} For $a<0$  the integration interval is $[a,0]$
    and $c<a$ or $c>0$ and $r$ has to be such that $a^r>0$.
\item [{\bf v)}] This case splits into two subcases:
    \subitem {\bf 1)}For $a>0$  the integration interval is $[0,a]$.
    \subitem {\bf 2)}For $a<0$  the integration interval is $[a,0]$
and $r$ has to be such that $a^r>0$.
\item [{\bf vi)}] This case splits into two subcases:
    \subitem {\bf 1)}For $a>0$  the integration interval is $[0,a]$.
    \subitem {\bf 2)}For $a<0$  the integration interval is $[a,0]$
    and $r$ have to be such that $a^r>0$.
\item [{\bf vii)}] In this case ${\cal R}_{AB}(q^n)$ are not positive for
 $n$ large anough.
\item [{\bf viii)}] In this case ${\cal R}_{AB}(q^n)=0$ and ${\cal D}_{AB}(q^n)=0$ for $n\in \Bbb N$.
\end{enumerate}
\end{proposition}

{\it Proof} :
\begin{enumerate}
\item[\bf i)] The equation $B(\omega)\varrho(\omega)=0$ is solved by $aq^{-k+1}$
and by $bq^{-k+1}$ for $k\in\Bbb N$. For any function $f(\omega)$
and any $k,l\in\Bbb N$, using (\ref{Cj}) and (\ref{Ck}) one can
obtain
\begin{equation}
 \int_{aq^{-l+1}}^{aq^{-k+1}}f(\omega)d\sigma_{AB}(\omega)=0= %
 \int_{bq^{-l+1}}^{bq^{-k+1}}f(\omega)d\sigma_{AB}(\omega)
\end{equation}
and
\begin{equation}
 \int_a^bf(\omega)d\sigma_{AB}(\omega)= %
 \int_{aq^{-k+1}}^{bq^{-l+1}}f(\omega)d\sigma_{AB}(\omega).
\end{equation}
Hence have the integration interval is $[a,b]$. The condition of
positivity of $d\sigma_{AB}(\omega)$
\begin{equation}
 \int_a^bf(\omega)d\sigma_{AB}(\omega)>0\;\;for f>0
\end{equation}
is equivalent to
\begin{equation}
  a\varrho(q^ia)<0\;\;and\;\;  b\varrho(q^ib)>0\;\;for
\;\;  i=0,1,\ldots.
 \end{equation}
 The continuity of $\varrho$ at $\omega=0$  gives $a<0<b$ and the
 inequalities
\begin{equation}
 \varrho(q^ia)>0\;,\;\; \; \varrho(q^ib)>0\;\;for\;\;  i=0,1,\ldots.
 \end{equation}
 which are solved by $\alpha$)-$\zeta$).\\
 The proofs of ii)-viii) are similar to the one above.  QED
  \end{enumerate}

  The above class of orthogonal polynomials, which we call
  the q-Hahn polynomials, contains, as a special cases the families
  of orthogonal polynomials well known from literature. Using a very good
  paper \cite{K-S} we obtain the following identification:
  \begin{enumerate}
  \item [\bf 1)]Putting in i)${\beta}$)  $\;d=1$ we have  the Big q-Jacobi
  polynomials. If in addition we put $b=q$ and $c=\frac{a}{q}$ we
  obtain the Big q-Legendre polynomials.
  \item [\bf 2)]Putting in ii) $b<1$ and $c=1$ we obtain the Big
  q-Laguerre  polynomials.
  \item [\bf 3)]Putting in iii) $b=1$ we obtain the
  Al-Salam-Carlitz I  polynomials. If in addition we assume $a=-1$ we
  obtain the Discrete q-Hermite I polynomials.
  \item [\bf 4)]Putting in iv)1) $a=1$ we obtain the
 Little q-Jacobi polynomials. If in addition we put $c=\frac{1}{q}$ %
   and $r=0$ we  obtain the Little q-Legendre polynomials.
  \item [\bf 5)]Putting in v)1) $a=1$ we obtain the
  Little q-Laguerre/Wall  polynomials.
  \item [\bf 6)]Putting in vi)1) $r=0$ we obtain the Alternative
q-Charlier  polynomials.
  \end{enumerate}

We will find now the equations for the moments
\begin{equation}
\mu_n=\int_a^b\omega^nd\sigma_{AB}\left(\omega\right)\label{Cw}
\end{equation}
 of the measure $d\sigma_{AB}$.\\%

 From Section 2 it is clear that once the moments are known one
 may determine many important physical characteristic of the
 system under consideration.

 Multiplying $q$-difference Pearson equation (\ref{Cf}) by
 $\omega^nq^n$ and using Lebnitz rule for $q$-derivative we obtain the
 following three-term recurrence equation
 \begin{equation}
-\left[n\right]\left(b_2\mu_{n+1}+b_1\mu_n+b_0\mu_{n-1}\right) =
q^n\left(a_1\mu_{n+1}+a_0\mu_n\right)\label{Cx}
\end{equation}
for $n\geq1$, and
\begin{equation}
a_1\mu_1+a_0\mu_0=0.\label{Cy}
\end{equation}
The initial rule $\mu_0=\int_a^bd\sigma_{AB}(\omega)$  for this
recurrence can be calculated in straightforward way. In terms of
the notation and classification introduced in {\bf Proposition
3.2} we have
\begin{enumerate}
\item [i)]
\begin{eqnarray}
\mu_0&=&(1-q)(b-a)\frac{(q;q)_\infty(q\frac{b}
{a};q)_\infty(q\frac{a}{b};q)_\infty(\frac{ab}{cd} ;q)_\infty}
{(\frac{a}{c};q)_\infty
(\frac{a}{d};q)_\infty(\frac{b}{c};q)_\infty (\frac{b}{d};
q)_\infty}
\end{eqnarray}
\item [ii)]
\begin{eqnarray}
\mu_0&=&(1-q)(b-a)\frac{(q;q)_\infty(q\frac{b}
{a};q)_\infty(q\frac{a}{b};q)_\infty} {(\frac{a}{c};q)_\infty
(\frac{b}{c};q)_\infty }
\end{eqnarray}
\item [iii)]
\begin{eqnarray}
\mu_0&=&(1-q)(b-a)(q;q)_\infty(q\frac{b}
{a};q)_\infty(q\frac{a}{b};q)_\infty
\end{eqnarray}
\item [iv)]
\begin{eqnarray}
\mu_0&=&(1-q)a^{r+1}\frac{(q;q)_r} {(\frac{a}{c}; q)_{r+1}}
\end{eqnarray}
\item [v)]
\begin{eqnarray}
\mu_0&=&(1-q)a^{r+1}(q;q)_r
\end{eqnarray}
\item [vi)]
\subitem a)
\begin{eqnarray}
\mu_0&=&(1-q)a^{r+1}\frac{(q;q)_\infty(-aq^{r+1};q)_\infty}
{(-a;q)_\infty(-\frac{q}{a}; q)_\infty}
\end{eqnarray}
\subitem b)
\begin{eqnarray}
\mu_0&=&(1-q)a^{r+1}\frac{(q;q)_\infty(aq^{r+1};q)_\infty}
{(a;q)_\infty(\frac{q}{a}; q)_\infty}\;\;.
\end{eqnarray}
\end{enumerate}
Let us note that replacing in iv)-vi) $r$  by $r+n,\;\;n\in \Bbb
N$ we obtain the moments $\mu_n$ for the corresponding cases.

In order to consider the cases i)-iii) let us introduce a real
function $\mu$ satisfying the equation
\begin{equation}
\left(1-\omega\right)B\left(Q\right)\mu\left(\omega\right)
+\left(1-q\right)\omega QA\left(Q\right)\mu\left(\omega\right) =
0. \label{Cz}
\end{equation}
It is easy to check that $ \mu\left(q^{n+1}\right) $ satisfies the
recurrence  equation (\ref{Cx}), hence  $ \mu(q^{n+1}) =\mu_n $.\\
 Reexpressing
(\ref{Cz}) in  the form
\begin{equation}
\left(\frac{B\left(Q\right)}{B\left(q^{-1}Q\right)-\left(1-q\right)
q^{-1}Q A
\left(q^{-1}Q\right)}-\omega\right)\mu\left(\omega\right)=0\label{Caa}
\end{equation}
and the equation (\ref{Cf}) in the form
\begin{equation}
\label{Cab}
\left(\frac{B\left(q\omega\right)}{B\left(\omega\right)-\left(1-q\right)
\omega A\left(\omega\right)}-Q\right)\varrho \left(\omega\right)=0
\end{equation}
one may observe some symmetry between the equation on $\varrho$
and the equation for the moment function $\mu$. After the
substitution of the form $Q\rightarrow q\omega$ and
$\omega\rightarrow Q$ the operator from (\ref{Caa}) transforms
into the one of (\ref{Cab}). The equation (\ref{Caa}) as well as
the equation (\ref{Cab}) can be easily solved.

For example, if we assume that $B(1)=0$ then  (\ref{Caa}) can be
written in the form
\begin{equation}
\label{Cac}\partial_{\cal R}\mu(\omega)=\mu(\omega),
\end{equation}
where $\partial_{\cal R}$ is $\cal R$-derivative. The function
$\cal R$ is  here given by
\begin{equation}
{\cal R}(\omega)=\frac{B(\omega)}{B(q^{-1} \omega) -(1-q)q^{-1}
\omega A(q^{-1} \omega)}.
\end{equation}
 Then one of the two linearly independent solutions of
(\ref{Caa}) is simply $\cal R$-exponential $Exp_{\cal R}$. In this
case it is given as the basic hypergeometric series
\begin{equation}
\mu_1\left(\omega\right)= Exp_{\cal
R}(\omega)={_3\Phi_2}\left(^{\frac{1}{a},\;1,\;q}_{\;\frac{1}{c},\;\frac{1}{d}};q;
\omega\right)
\end{equation}
where $a\neq1$ is the root of the polynomial $B(\omega)$ and
$c,\;d$  are roots of the polynomial
$B(q^{-1}\omega)-(1-q)q^{-1}\omega A(q^{-1}\omega)$. The function
${_3\Phi_2}$ is defined in \cite{G-R}, \cite{K-S}. \\
 The second solution $\mu_2\left(\omega\right)$ is related to
$\mu_1\left(\omega\right)$ by the following formula ($q$-version
of Wronskian):
\begin{equation}\label{Cad}
\mu_2\left(\omega\right)\mu_1\left(q\omega\right)-\mu_2
\left(q\omega\right)\mu_1\left(\omega\right)=x^\lambda
\frac{\left(\alpha
q;q\right)_\infty}{\left(\omega;q\right)_\infty}
\end{equation}
where $q^\lambda=\frac{b_2}{b_0},\;\; \alpha=
\frac{(1-q)a_1-b_2}{b_0}$.

Any solution $\mu\left(\omega\right)$ is a linear combination of
$\mu_1\left(\omega\right)$ and $\mu_2\left(\omega\right)$. We are
then getting the following formulae for moments
\begin{equation}
\mu_n=\mu\left(q^{n+1}\right)=c_1\mu_1\left(q^{n+1} \right)
+c_2\mu_2\left(q^{n+1}\right)
\end{equation}
where the constants $c_1,c_2$ are determined by
 \begin{eqnarray}
 &&\;\;\;\;\;\;\;\mu_0=\int_a^bd\sigma_{AB}=c_1\mu_1\left(q\right)+c_2
 \mu_2\left(q\right)\\
 &&a_0\left(c_1\mu_1\left(q\right)+c_2\mu_2\left(q\right)\right)
 +a_1\left(c_1\mu_1\left(q^2\right)+c_2\mu_2\left(q^2\right)\right)=0.
 \end{eqnarray}
\section*{ Appendix A
\\ The affine difference calculus and q-Hahn's orthogonal polynomials}
\renewcommand{\theequation}{A.\arabic{equation}}
\setcounter{equation}{0}

In this section we present the preliminary considerations related
to the calculus generated by the action of the affine group
$A_{+}$ on the real line. Let us define the linear representation
of $A_{+}$
\begin{eqnarray}
\left( \frak{L}_{q,h}\varphi \right) \left( x\right) :=\varphi
\left( qx+h\right),\;\;\left( q,h\right) \in A_{+}=\left\{ \left(
q,h\right) :\;q>0,\;\;h\in \Bbb{R} \right\}   \label{A.a}
\end{eqnarray}
acting on the functions  $\varphi $  from the algebra $\frak{F}$.
Since our consideration will be formal in its character we do not
impose any additional conditions on $\frak{F}$.

According to \cite{H}  we introduce the derivative operator
\begin{equation}
\left( \partial _{q,h}\varphi \right) \left( x\right)
:=\frac{\varphi \left( x\right) -\varphi \left( qx+h\right)
}{x-\left( qx+h\right) }  \label{A.b}
\end{equation}
as a natural generalization of the q- derivative $\partial
_q:=\partial _{q,0}$ and of the difference derivative $\partial
_h:=\partial _{1,h}$.

The Leibnitz rule for the derivative $\partial _{q,h}$ is
\begin{equation}
\left( \partial _{q,h}\varphi \psi \right) \left( x\right) =\left(
\partial
_{q,h}\varphi \right) \left( x\right) \psi \left( x\right) +\left( \frak{L}%
_{q,h}\varphi \right) \left( x\right) \left( \partial _{q,h}\psi
\right) \left( x\right)  \label{A.c}
\end{equation}
There is also the following equivariance property:
\begin{equation}
\frak{L}_{c,t}^{-1}\circ \partial _{q,h}\circ
\frak{L}_{c,t}=c\,\partial _{q,ch+\left( 1-q\right) t} \label{A.d}
\end{equation}
and enables us to reduce $\left( q,h\right) -$analysis to
$q-$analysis.  We have for example
\begin{equation}
\partial _{q,h}=\frak{L}_{1,\frac h{1-q}}^{-1}\circ \partial _q\circ \frak{L}%
_{1,\frac h{1-q}}.  \label{A.e}
\end{equation}

Let us now solve  the equation
\begin{equation}
\partial _{q,h}\varphi =\varrho  \label{A.f}
\end{equation}
for the given function $\varrho \in \frak{F}$.\quad In the order
to do this, we apply the operator $\frak{L}_{q,h}^k$\quad to
(\ref{A.f}), and we find that
\begin{equation}
\frak{L}_{q,h}^k\varphi \left( x\right)
-\frak{L}_{q,h}^{k+1}\varphi \left( x\right) =q^k\left[ \left(
1-q\right) x-h\right] \frak{L}_{q,h}^k\varrho \left( x\right) .
\label{A.g}
\end{equation}

Summing up both sides of the identity\quad (\ref{A.g})\quad with respect to $%
k$\quad we get
\begin{equation}
\varphi \left( x\right) -\varphi \left( x^\infty \right) =
\sum_{k=0} ^\infty \left[ x-\left( qx+h\right) \right] q^k\varrho
\left( q^kx+\frac{1-q^k}{1-q}h\right) \label{A.h}
\end{equation}
where
\begin{equation}
x^\infty = \lim_{k\rightarrow \infty } \left(
q^kx+\frac{1-q^k}{1-q}h\right)=\frac{h}{1-q}.\label{A.i}
\end{equation}
The equation (\ref{A.h}), justifies the following definition of the $\left( q,h\right) -$%
integral
\begin{equation}
\int_{q,h}\varrho \left( x\right) =\int_{x^\infty }^{x} \varrho
\left( t\right) d_{q,h}t:=\sum^ \infty _{k=0}
\left[ x-\left( qx+h\right) \right] q^k\varrho \left( q^kx+\frac{1-q^k}{1-q}%
h\right).  \label{A.j}
\end{equation}

The $\left( q,h\right) -$integral operator is the right inverse of the $%
\left( q,h\right) -$derivative operator
\begin{equation}
\partial _{q,h}\circ \int_{q,h}=id,  \label{A.k}
\end{equation}
and, moreover,
\begin{equation}
\int_{q,h}\circ \partial _{q,h}=id-\delta _{1,\infty }.
\label{A.l}
\end{equation}

The oprerator\quad $\delta _{1,\infty }$\quad is an idempotent
operator defined by
\begin{equation}
\left( \delta _{1,\infty }\varphi \right) \left( x\right) =\varphi
\left( x^\infty \right),  \label{A.m}
\end{equation}
projecting the function on the constants.

Like in (\ref{A.e}) we have
\begin{equation}
\int_{q,h}=\frak{L}_{1,\frac h{1-q}}^{-1}\circ \int_q\circ
\frak{L}_{1,\frac h{1-q}},  \label{A.n}
\end{equation}
which reduces (by the translation authomorfism\quad
$\frak{L}_{1,\frac
h{q-1}}$)\quad the $\left( q,h\right) -$integral to the Jackson integral $%
\int_q=\int_{q,0}$.\ The integration on the interval $\left[ a,b\right] $%
\quad can be defined by
\begin{equation}
\int_a^b\varrho \left( t\right) d_{q,h}t= \int_{b_\infty }^b
\varrho \left( t\right) d_{q,h}t- \int_{a_\infty }^a\varrho \left(
t\right) d_{q,h}t . \label{A.o}
\end{equation}

If  $q\rightarrow 1$, the calculus presented above corresponds to
the difference calculus. For $h\rightarrow 0$  one obtains
$q-$difference calculus. The differential calculus will be
obtained when  $q\rightarrow 1$,\ $h\rightarrow 0$.

Let us mention finally that the identities (\ref{A.e}) and
(\ref{A.n}) enable us to reduce $\left( q,h\right) $-calculations
to the $q$-calculations. This property motivates us to discuss the
case of the $q-$ analysis in this paper.

\section*{ Appendix B
\\ Proof of the Theorem 3.2}
\renewcommand{\theequation}{B.\arabic{equation}}
\setcounter{equation}{0}

$A\Leftrightarrow B$

The monic OPS $\left\{ P_n\left( \omega\right) \right\}
_{n=0}^\infty $ is uniquely defined by the weight function
$\varrho \left( \omega\right) $ and the interval
$\left[a,b\right]$. In order to prove the equivalence of the
properties $A$  and  $B$ \quad it is sufficient to show that the
system of polynomials defined by ( \ref{Cn}) is a monic OPS. In
order to do that let us reexpress the function
\begin{equation}
F_k\left( \omega;n\right) :=\partial _q^k\left[ \varrho \left(
\omega\right) B\left( \omega\right) B\left( q^{-1}\omega\right)
\ldots B\left( q^{-\left( n-1\right) }\omega\right) \right]
\label{B.o}
\end{equation}
in the following way
\begin{equation}
F_k\left( \omega;n\right) =\varrho \left( \omega\right) B\left(
\omega\right) B\left( q^{-1}\omega\right) \ldots B\left(
q^{-\left( n-1-k\right) }\omega\right) R_{k,n}\left(
\omega\right)\;\;\;k=0,1,\ldots,n-1 \label{B.p}
\end{equation}
\begin{equation}
F_n\left( \omega;n\right) =\varrho\left( \omega\right)
R_{n,n}\left( \omega\right)
\end{equation}
where $R_{k,n}\left( \omega\right) $ is a polynomial of degree not
greater than $k\leq n$. These polynomials do and satisfy the
recurrence formula
\begin{equation}
R_{k+1,n}\left( \omega\right) =A\left( \omega\right) R_{k,n}\left(
q\omega\right) +B\left(
q^{-\left( n-1-k\right) }\omega\right) \partial _qR_{k,n}\left( \omega\right) +\frac{%
B\left( q^{-\left( n-1-k\right) }\omega\right) -B\left(
\omega\right) }{\left( 1-q\right) \omega}R_{k,n}\left(
q\omega\right) , \label{B.r}
\end{equation}
for $k=0,1,\ldots,n-1$ with the initial condition
$R_{0,n}\left(\omega\right) \equiv 1$.

For $k<n$, applying (\ref{B.p}) we have:
\begin{eqnarray}
&&\int_a^bq^{\frac{k\left( k+1\right) }
2}\omega^k\widetilde{P}_n\left( \omega\right) \varrho \left(
\omega\right) d_q\omega =c_n \int_a^bq^{\frac{k\left( k+1\right)
}2}\omega^k\partial
_q^nF_0\left( \omega;n\right) d_q\omega= \nonumber \\%
 &&\;\;\;\;
=\left. c_nq^{\frac{k\left( k+1\right) }2}\omega^kF_{k-1}\left(
\omega;n\right)\right| _a^b-c_n\left[ k\right] \int_a^b q^{\frac{
k\left( k+1\right) }2}\omega^{k-1}\partial _q^{n-1}F_0\left(
\omega;n\right) d_q\omega= \nonumber \\%
&&\;\;\;\;=\left. c_nq^{\frac{k\left( k+1\right)
}2}\omega^k\varrho \left( \omega\right) B\left( \omega\right)
B\left( q^{-1}\omega\right) \ldots B\left( q^{-\left( n-1-\left(
k-1\right) \right) }\omega\right) R_{k-1,n}\left( \omega\right)
\right| _a^b- \label{B.s}\\%
&&\;\;\;\;\;\;\;\;-c_n\left[ k\right] \int_a^bq^{\frac{k\left(
k+1\right) }2}\omega^{k-1}\partial _q^{n-1}F_0\left(
\omega;n\right) d_q\omega=\;\;\ldots \;\;=\nonumber \\%
&&\;\;\;\; =\left( -1\right) ^k\left[ k\right] \left[ k-1\right]
\ldots \left[ 1\right] c_n \int_a^b\partial
_q^{n-k}F_0\left( \omega;n\right) d_q\omega= \nonumber\\%
&&\;\;\;\; =\left. \left( -1\right) ^k\left[ 1\right] \ldots
\left[ k\right] \varrho \left( \omega\right) B\left( \omega\right)
B\left( q^{-1}\omega\right) \ldots B\left( q^{-\left( n-1-\left(
k-1\right) \right) }\omega\right) \right| _a^b=\nonumber\\%
&&\;\;\;\; =0.\nonumber
\end{eqnarray}
This shows, that polynomials $\widetilde{P}_n\left( \omega\right) $%
\ , $n\in \Bbb{N\cup }\left\{ 0\right\} $, form an OPS. By the
proper choice of the normalizing constants  $c_n$ one can obtain
the monic OPS.

$B\Rightarrow C$

We have proved the validity of the $q-$ Rodrigues formula for any
Pearson data. So by {\bf Theorem 3.1}, we have
\begin{equation}
\frac 1{\left[ n\right] }\partial _q\widetilde{P}_n\left(
\omega\right) =c_{n-1}^{\left(1\right)}\frac 1{\varrho
^{\left(1\right)}\left( \omega\right) }\partial _q^{n-1}\left[
\varrho ^{\left(1\right)}\left( \omega\right) B^{\left( 1\right)
}\left( \omega\right) B^{\left( 1\right) }\left(
q^{-1}\omega\right) ...B^{\left( 1\right) }\left( q^{-\left(
n-2\right) }\omega\right) \right] \label{B.t}
\end{equation}
for the Pearson data \quad $\left( A^{\left( 1\right) }\left(
\omega\right) ,B^{\left( 1\right) }\left( \omega\right) \right)
$\quad given by (\ref{Ch}-\ref{Ci}). Using now the equality
\begin{equation}
Q\varrho \left( \omega\right) B\left( \omega\right) ...B\left(
q^{-\left( n-1\right) }\omega\right) =\varrho
^{\left(1\right)}\left( \omega\right) B^{\left( 1\right) }\left(
\omega\right) B^{\left( 1\right) }\left( q^{-1}\omega\right)
...B^{\left( 1\right) }\left( q^{-\left( n-2\right) }\omega\right)
\label{B.u}
\end{equation}
and substituting (\ref{B.t})  into (\ref{Cn})  we find
\begin{equation}
\widetilde{P}_n\left( \omega\right) =c_n\frac 1{\varrho \left(
\omega\right) }\partial _q^n\left[ Q^{-1}\varrho
^{\left(1\right)}\left( \omega\right) B^{\left( 1\right) }\left(
\omega\right) B^{\left( 1\right) }\left( q^{-1}\omega\right)
...B^{\left( 1\right) }\left( q^{-\left( n-2\right) }\omega\right)
\right] =  \label{B.w}
\end{equation}
\begin{eqnarray*}
\ &=&c_nq^{-\left( n-1\right) }\frac 1{\varrho \left(
\omega\right) }\partial _qQ^{-1}\partial _q^{n-1}\left[ \varrho
^{\left(1\right)}\left( \omega\right) B^{\left( 1\right) }\left(
\omega\right) B^{\left( 1\right) }\left( q^{-1}\omega\right)
...B^{\left( 1\right) }\left( q^{-\left( n-2\right) }\omega\right)
\right] = \\ \ &=&c_nq^{-\left( n-1\right) }\frac 1{\varrho \left(
\omega\right) }\partial
_qQ^{-1}\frac{\left[ n\right] \varrho ^{\left(1\right)}\left( \omega\right) }{%
c_{n-1}^{\left(1\right)}}\partial _q\widetilde{P}_n\left(
\omega\right) = \\ \ &=&\frac{c_n}{c_{n-1}^{\left(1\right)}}
q^{-\left( n-1\right) }\left[ n\right] \frac 1{\varrho \left(
\omega\right) }\partial _q\varrho \left( \omega\right) B\left(
\omega\right) Q^{-1}\partial _q\widetilde{P}_n\left( \omega\right)
= \\ \ &=&\frac{c_n}{c_{n-1}^{\left(1\right)}}q^{-\left(
n-1\right) }\left[ n\right] \left[ A\left( \omega\right)
\partial _q\widetilde{P}_n\left( \omega\right) +B\left( \omega\right)
\partial _qQ^{-1}\partial _q\widetilde{P}_n\left( \omega\right) \right] .
\end{eqnarray*}
We have proved (\ref{Co}). In order to show (\ref{Cp}) we compare
the coefficients of $x^n$ on both sides of (\ref{Co}).
Additionally we find the formula
\begin{equation}
\frac{c_{n-1}^{\left(1\right)}}{c_n}=\left[ n\right] q^{-\left(
n-1\right) }\lambda _n  \label{B.x}
\end{equation}
for the normalizing coefficients. Later we will use (\ref{B.x})
for the calculation of $c_n$.

$C\Rightarrow B$

The proof goes by the induction. It is easy to see that for $n=1$
 (\ref{Cn})follows from  (\ref{Co}). Let us assume that it is true for $n-1$.
We prove it for $n$:

From our assumption and {\bf Theorem 3.1} we have
\begin{equation}
\frac{1}{\left[n\right]}\partial_q\widetilde{P}_n\left(\omega
\right)= c_{n-1}^{\left(1\right)}\frac{1}{\varrho^{\left(1\right)}
\left(\omega\right)}\partial_q^{n-1}\left(\varrho^{
\left(1\right)}  \left(\omega\right)B\left(\omega\right)\ldots B
\left(q^{-\left(n-2\right)}\omega\right)\right).
\end{equation}
Using (\ref{Cn}) and the {\bf Proposition 3.1}, we obtain thesis
after a simple calculations.

$C \Leftrightarrow D$

From (\ref{Cp}) we have
\begin{equation}
\widetilde{P}_n\left( \omega\right) =\frac{\left[ n\right]
}{\lambda _n}\left( A\left( \omega\right) +B\left( \omega\right)
\partial _qQ^{-1}\right) \frac 1{\left[ n\right] }\partial
_q\widetilde{P}_n\left( \omega\right) . \label{C.f}
\end{equation}
According to the {\bf Theorem 3.1},  the polynomials $\left\{
\frac 1{\left[ n\right] }\partial _q\widetilde{P}_n\left(
\omega\right) \right\} _{n=0}^\infty $  form the monic OPS with
respect to the data\quad $\left( A^{\left( 1\right) }\left(
\omega\right) ,B^{\left( 1\right) }\left( \omega\right) \right) $
given by (\ref{Ch}-\ref{Ci}). We can thus apply the formula
(\ref{C.f}) with $ \left( A^{\left( 1\right) }\left( \omega\right)
,B^{\left( 1\right) }\left( \omega\right) \right) $ to the
polynomial $\frac 1{\left[ n\right] }\partial _qP_n\left(
\omega\right) $. Repeating this procedure $n$-times and using the
formula
\begin{equation}
\lambda _n^{\left( k\right) }=a_1^{\left( k\right) }\left[
n\right] +b_2\left[ n\right] \left[ n-1\right] q^{-\left(
n-1\right) }  \label{C.g}
\end{equation}
where $a_1^{\left( k\right) }$ is defined by\ (\ref{Cs}), we
obtain\ (\ref{Cr}).

$B\Rightarrow E$

In order to prove that the recurrence formula (\ref{Ct}) holds we
use the identity
\begin{equation}
\partial _q^k\left[ \varrho \left( \omega\right) B\left( \omega\right) B\left(
q^{-1}\omega\right) ...B\left( q^{-\left( n-1\right)
}\omega\right) \right] =\varrho \left( \omega\right) B\left(
\omega\right) B\left( q^{-1}\omega\right) ...B\left( q^{-\left(
n-1-k\right) }\omega\right) R_{k,n}\left( \omega\right)
\label{C.h}
\end{equation}
where the polynomial $R_{k,n}\left( \omega\right) $ satisfies the
recurrence equation (\ref{B.r}). From the eq. (\ref{C.h}) and
Rodrigues formula one has
\begin{equation}
\widetilde{P}_n\left( \omega\right) =c_nR_{n,n}\left(
\omega\right) . \label{C.i}
\end{equation}
Let us denote by\ $\alpha _k$,\ $\beta _k$ and $\gamma _k$ the
three highest coefficients of the polynomial
\begin{equation}
R_{k,n}\left( \omega\right) =\alpha _k\omega^k+\beta
_k\omega^{k-1}+\gamma _k\omega^{k-2}+...\;\;.  \label{C.j}
\end{equation}
After substituting (\ref{C.j}) into (\ref{C.i}) and comparing the
coefficients of the monomials $\omega^{k+1}$,\ $\omega^k$\ and\
$\omega^{k-1}$\ we obtain the following system of the recurrence
equations
\begin{eqnarray}
\alpha _{k+1} &=&\left( a_1-b_2\left[ -2n+2+k\right] \right)
q^k\alpha _k \nonumber  \label{C.k} \\ \beta _{k+1} &=&\left(
a_0-b_1\left[ -n+1\right] \right) q^k\alpha _k+\left(
a_1-b_2\left[ -2n+3+k\right] \right) q^{k-1}\beta _k  \label{C.kk}
\\ \gamma _{k+1} &=&b_0\left[ k\right] \alpha _k+\left(
a_0-b_1\left[ -n+2\right] \right) q^{k-1}\beta _k+\left(
a_1-b_2\left[ -2n+4+k\right] \right) q^{k-2}\gamma _k. \nonumber
\end{eqnarray}
One can solve them by iteration:
\begin{equation}
\alpha _k=q^{\frac{k\left( k-1\right) }2}\prod_{l=0}^k-1 \left(
a_1-b_2\left[ -2n+2+l\right] \right)  \label{C.l}
\end{equation}
\begin{equation}
\beta _k=\frac{\left[ n\right] }{q^{n-1}}\cdot \frac{a_0-b_1\left[
-n+1\right] }{a_1-b_2\left[ -2n+2\right] }\cdot \alpha _k
\label{C.m}
\end{equation}
\begin{equation}
\gamma _k=\frac{\left( 1-q^n\right) \left( 1-q^{n-1}\right)
}{\left( 1-q\right) ^2\left( 1+q\right) }\cdot \frac{\left(
a_0-b_1\left[ -n+2\right] \right) \left( a_0-b_1\left[ -n+1\right]
\right) +b_0\left( a_1-b_2\left[ -2n+2\right] \right) }{\left(
a_1-b_2\left[ -2n+2\right] \right) \left( a_1-b_2\left[
-2n+3\right] \right) }\alpha _k  \label{C.n}
\end{equation}
Thus for the monic polynomial
\begin{equation}
\widetilde{P}_n\left( \omega\right) =\omega^n+\beta \left(
q^n\right) \omega^{n-1}+\gamma \left( q^n\right) \omega^{n-2}+...
\label{C.o}
\end{equation}
we find that the coefficients
\begin{eqnarray*}
\beta \left( q^n\right) &:=&\frac{\beta _n}{\alpha _n} \\ \gamma
\left( q^n\right) &:=&\frac{\gamma _n}{\alpha _n}
\end{eqnarray*}
are given by the rational functions (\ref{Cc}-\ref{Ccc}). Using
the three terms recurrence relation (\ref{Cu}) and (\ref{C.j}) we
obtain the formula (\ref{Ca}-\ref{Cb}) for the structural
functions\ $\frak{R}$\ and\ $\frak{D}$\ .

$E \Rightarrow A$

The recurrence formula (\ref{Ct}) rewriten for the orthonormal
polynomials $\left\{P_n\left(\omega\right)\right\}_{n=0}^\infty$
takes the form  (\ref{By}) which means that the Hamiltonian
$H_{red}$ (\ref{Br}) has the Jacobi matrix of the type $D$. Thus
(see \cite{A}) this Hamiltonian is essentially selfadjoint and has
simple spectrum. This shows that there is a unique measure
$d\sigma_{AB}$ such that
\begin{equation}
\int_a^bP_n\left(\omega\right)P_m\left(\omega\right)d\sigma_{AB}
\left(\omega\right)=\delta_{nm}.
\end{equation}

\section*{ Acknowledgements}
\renewcommand{\theequation}{A.\arabic{equation}}
\setcounter{equation}{0} Two of the authors (A.O., A.T.) would
like to thanks M.Rahman for his interest to the subject and for
discussions on orthogonal polynomials. The discussions with Z.
Hasiewicz on possible applications were also important.



\begin{thebibliography}{99}
\bibitem[A]{A}N.I. Ahiezer, " The classical moment problem",
Hafner Publ. Co., N.Y., 1965
\bibitem[A-G]{A-G}N.I. Ahiezer, I.M. Glazman, "Theory of linear operators in Hilbert space", Nauka,
Moscow, 1966 (in Russian)
\bibitem[A-I]{A-I} V.A. Andreev, O.A. Ivanova, ''The dynamics of
three-boson interaction and algebraic Bethe ansatz'', Phys. Lett.
A 171 (1992) 145-150
\bibitem[Ch]{Ch}  T.S.Chihara ''An introduction to orthogonal
polynomials''Gordon and Breach, New York, 1978
\bibitem[G-R]{G-R}  G. Gasper, M. Rahman ''Basic hypergeometric series''
Cambridge Un. Press, 35 (1990);
\bibitem[H]{H}  W. Hahn, ''\"Uber Orthogonalpolynome die
q-Differenzengleichungen gen\"ugen'' Math.Nachr., 2 (1949), 4-34;
\bibitem[J1]{J1}  B. Jur\v{c}o ''Classical Yang-Baxter
equations and quantum integrable systems'' J. Math. Phys.
\underline{30} (6) June 1989
\bibitem[J2]{J2}  B. Jur\v{c}o ''On quantum integrable
models releted to nonlinear quantum optics. An algebraic Bethe
ansatz approach'' J. Math. Phys. \underline{30} (8) August 1989
\bibitem[Kar1]{Kar1} V.P. Karassiov ''Polynomial Lie algebras and associated
pseudogroup structures in composite quantum models'' Rep. on Math.
Phys. vol. 40 (1997) N\underline{o}.2
\bibitem[Kar2]{Kar2} V.P. Karassiov ''sl(2) variational schems
for solving one class of nonlinear quantum models" Physics Letters
A 238 (1998) 19-28
\bibitem[K-S]{K-S} R. Koekoek, R.F. Swarttouw "The Askey-scheme
of hypergeometric orthogonal polynomials and its q-analogue"
Report no. 98-17, TUDelft (1998),
http://aw.twi.tudelft.nl/~koekoek/askey.html
\bibitem[M-G]{M-G} B.R. Mollow, R.J. Glauber "Quantum Theory of
Parametric Amplification" in "Nonclassical Effacts in Quantum
Optics", Amer. Inst. of Phys., New York, 1991
\bibitem[Odz]{Odz} A. Odzijewicz ''Quantum Algebras and q-Special Functins
Related to Coherent States Maps of the Disc'' Commun. Math. Phys.
192, 183-215 (1998)
\bibitem[R-S]{R-S}M. Reed, B. Simon "Methods of modern mathematical physics" Academic Press N.Y. London, 1972
\bibitem[St]{St} M. Stone "Linear Transformations in Hilbert Space and their Applications to
Analysis"N.Y., 1963
\bibitem[Su]{Su}  I.K. Sujetin ''Klassical orthogonal polynomials''
Moskwa ''Nauka'', 1979(in Russian)
\end{thebibliography}
\end{document}